\documentclass{IEEEoj}
\usepackage{cite}
\usepackage{amsmath,amssymb,amsfonts}
\usepackage{algorithmic}
\usepackage{graphicx,color}
\usepackage{textcomp}
\usepackage{cite}
\usepackage{amsmath,amssymb,amsfonts}
\usepackage{algorithmic}
\usepackage{graphicx,color}
\usepackage{textcomp}
\usepackage{booktabs}
\usepackage{xcolor,comment}
\usepackage{amsmath,amssymb,amsfonts}
\usepackage{algorithmic}
\usepackage{graphicx,makecell}
\usepackage{cite}
\usepackage{textcomp}
\usepackage{acronym}
\usepackage{array,url}
\usepackage{lscape,arydshln}
\def\BibTeX{{\rm B\kern-.05em{\sc i\kern-.025em b}\kern-.08em
    T\kern-.1667em\lower.7ex\hbox{E}\kern-.125emX}}
\AtBeginDocument{\definecolor{ojcolor}{cmyk}{0.93,0.59,0.15,0.02}}
\def\OJlogo{\vspace{-14pt}\includegraphics[height=28pt]{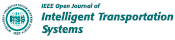}}
\begin{document}
\receiveddate{XX Month, XXXX}
\reviseddate{XX Month, XXXX}
\accepteddate{XX Month, XXXX}
\publisheddate{XX Month, XXXX}
\currentdate{XX Month, XXXX}
\doiinfo{OJITS.2022.1234567}

\title{XLM for Autonomous Driving Systems:\\A Comprehensive Review
}

\author{$^1$Sonda Fourati, $^2$Wael Jaafar, $^1$Noura Baccar, and $^3$Safwan Alfattani 
}
\affil{Mediterranean Institute of Technology (MedTech), Tunis, Tunisia.}
\affil{Software and IT Engineering Department, École de Technologie Supérieure (ÉTS), University of Quebec, Montreal, Canada.}
\affil{King AbdulAziz University (KAU), Rabigh, Saudi Arabia.}
\corresp{CORRESPONDING AUTHOR: Wael Jaafar (e-mail: wael.jaafar@etsmtl.ca).}


\markboth{Preparation of Papers for IEEE OPEN JOURNALS}{Author \textit{et al.}}

\begin{abstract}
Large Language Models (LLMs) have showcased remarkable proficiency in various information-processing tasks. These tasks span from extracting data and summarizing literature to generating content, predictive modeling,  decision-making, and system controls. Moreover, Vision Large Models (VLMs) and Multimodal LLMs (MLLMs), which represent the next generation of language models, a.k.a., XLMs, can combine and integrate many data modalities with the strength of language understanding, thus advancing several information-based systems, such as Autonomous Driving Systems (ADS). Indeed, by combining language communication with multimodal sensory inputs, e.g., panoramic images and LiDAR or radar data, accurate driving actions can be taken. In this context, we provide in this survey paper a comprehensive overview of the potential of XLMs towards achieving autonomous driving. Specifically, we review the relevant literature on ADS and XLMs, including their architectures, tools, and frameworks. Then, we detail the proposed approaches to deploy XLMs for autonomous driving solutions. Finally, we provide the related challenges to XLM deployment for ADS and point to future research directions aiming to enable XLM adoption in future ADS frameworks. 
\end{abstract}



\maketitle

\section{INTRODUCTION}
\subsection{General Context}
\IEEEPARstart{I}{n} today's dynamic and constantly evolving world, the public sector encounters numerous obstacles, particularly in ensuring safety and saving resources. In efforts to make roads safer, reduce human driving errors, increase mobility for those unable to drive, and enhance transportation efficiency, Autonomous Driving (AD), a.k.a., self-driving cars, is transforming our ways of traveling \cite{Zhu2019}. However, achieving fully autonomous driving faces several challenges such as unreliable perception and decision-making in complex and unpredictable traffic scenarios. 

To navigate these challenges, governments and public organizations are leveraging novel technologies, such as Generative Artificial Intelligence (GAI) \cite{cao2023comprehensive}. The latter stands out as a potential solution across multiple disciplines, such as public safety \cite{ooi2023potential,mahor2023iot,anderljung2023frontier}. GAI techniques, such as Large Language Models (LLMs) utilize advanced algorithms to generate human-like text or content based on input prompts or patterns in the data they trained on \cite{chang2024survey ,hadi2023survey}.
LLMs are initially trained on a large dataset of text data to learn semantics and general language patterns. This pre-training process involves tasks like predicting the next word in a sentence (language modeling) or filling in missing words based on the surrounding context. Embedding of customized data into LLMs allows for utilizing the strengths of pre-trained language models while adapting them to specific applications or domains \cite{cui2024survey, ge2023development, chen2023enhancing, wu2024accidentgpt}. 
For instance, there are biomedical LLMs such as Med-PALM \cite{qian2024liver, singhal2023large}, BioGPT \cite{ouis2024chestbiox}, BioBERT \cite{kafikang2023drug}, ClinicalBERT \cite{aden2024international}, which are fine-tuned to perform specialized tasks. 

Despite the LLMs' strong reasoning performance on several Natural Language Processing (NLP) tasks \cite{huang2024leveraging}, they are ``blind" to visuals since they are limited to understanding discrete text. In contrast, Vision Language Models (VLMs) can see well but are typically weaker in reasoning compared to LLMs \cite{wang2023review, maaz2023video, bai2023sequential,  xu2023lvlm,  zhao2024evaluating, li2024eyes}. 
Nevertheless, the complementarity of VLMs and LLMs paves the way towards a new field known as multimodal LLM \cite{zhang2024mm, yin2023survey}. 
Indeed, MLLMs can combine image, video, and audio data with the advanced reasoning capabilities of LLMs, thus they are well-equipped to execute a wide range of tasks, such as speech recognition, image classification, and text-to-video matching. We abbreviate the aforementioned language models as XLMs.  



The study of XLM for the advancement of autonomous driving started recently and is still in its infancy. For instance, authors of \cite{cui2024survey} highlighted the advantages of MLLM for AD and overviewed related works recently published. 
Also, authors in \cite{zhang2024trafficgpt, zheng2023trafficsafetygpt} introduced traffic safety decision-making through multi-modality representation learning.
They suggested that MLLM can potentially contribute to various aspects of traffic safety research. 
\subsection{Motivations, Objectives, and Contributions}
The rapid progress of XLMs has opened new frontiers in AD. Indeed, LLMs have shown strong abilities in understanding context, handling complex tasks, and generating answers. Moreover, their integration with multimodal data, to build MLLM systems, can enhance the system’s generalization and adapt to new scenarios with complex driving behaviors. With VLMs, the perception issue can be tackled and the decision-making process becomes more transparent.

Since 2023, several researchers have been investigating AD with LLM, VLM, or MLLM. However, to the best of our knowledge, there is no comprehensive work that gathered state-of-the-art works focusing on using LLM, VLM, and MLLM in AD together, nor did they evaluate the practical deployment of XLMs for AD. Hence, the main objective of this survey is to draw the full picture of the state-of-the-art regarding the use of XLMs for AD Systems (ADS) and to bridge the gap between their theory and practical use in ADS. Specifically, the contributions of this survey can be listed as follows:
\begin{itemize}
    \item Overview of the basic concepts of ADS and XLMs.
    \item Comprehensive overview of XLMs applied to ADS.

    \item Extensive study on recent datasets, tools, frameworks, and benchmarks that enable the practical implementation of XLMs in ADS.

    \item Identification of related challenges and future research directions to promote ADS deployment assisted with XLMs.    
\end{itemize}



\subsection{Organization of the Paper}
\begin{figure}
   \centering
   \includegraphics[trim={1cm 1.2cm 1.8cm 1cm},clip,width = 0.99\columnwidth] {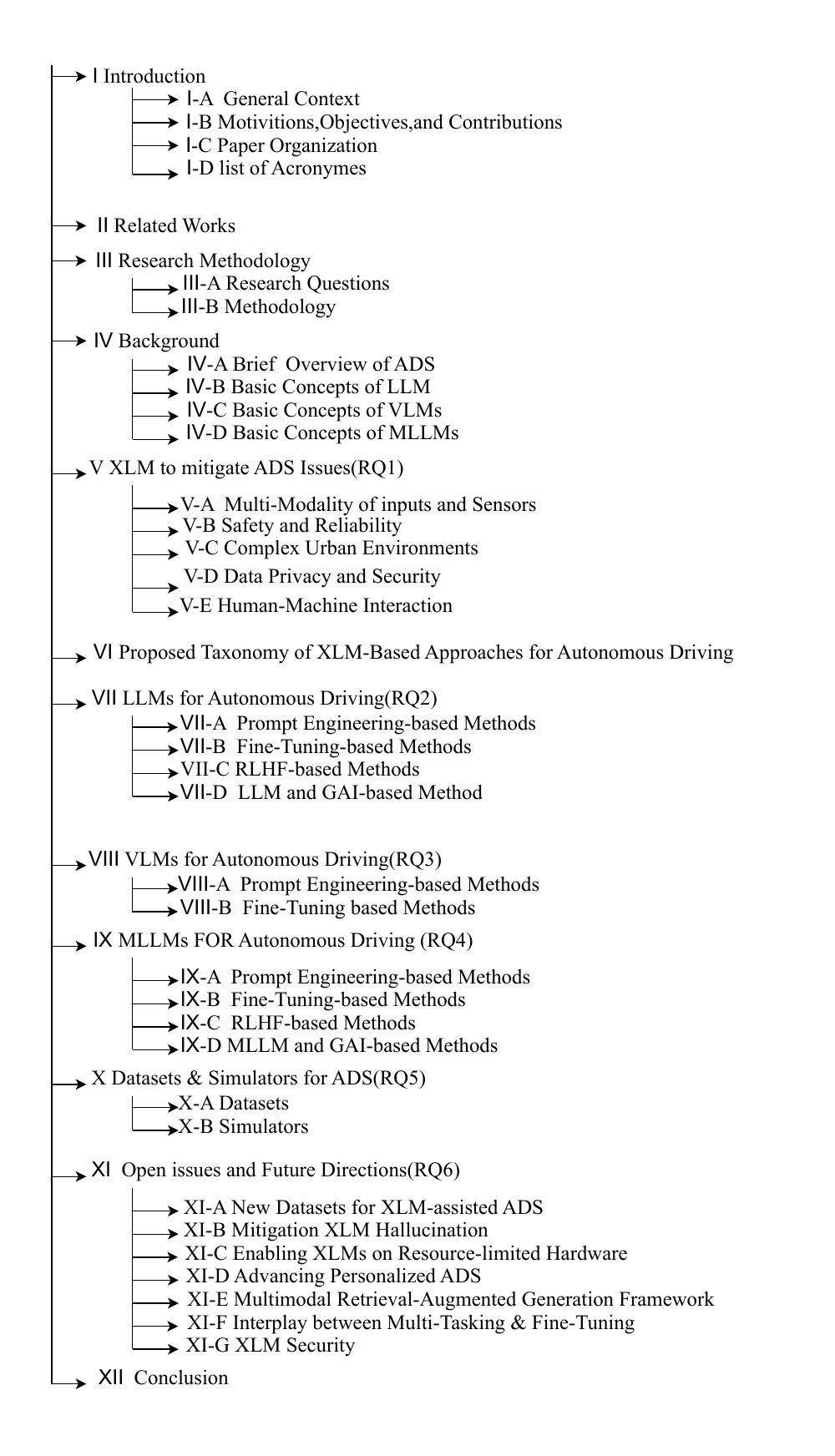}
  \caption{Paper organization.}
   \label{roadmap}
\end{figure}

The remaining of the survey is structured as follows. Section II summarizes related surveys that studied XLMs, a.k.a., Foundation Models (FM), used to achieve AD. Section III presents our research methodology and identifies the research questions of XLMs towards AD. Section IV introduces the background knowledge related to ADS and XLMs. Section V highlights the role of XLM to mitigate ADs challenges. Section VI presents the proposed taxonomy. Sections VII, VIII, and IX explore the technological advancements and recent works on the use of LLMs, VLMs, and MLLMs, respectively, towards AD. Section X assesses datasets and benchmarks that could be utilized to enable the practical deployment and testing of XLMs in ADS. Section XI discusses open issues and future research directions. Finally, section XII closes the survey. This organization is illustrated in Fig. \ref{roadmap}.

\subsection{List of Acronyms}
Below, we present the list of acronyms used in the survey paper. 
\begin{acronym}[MV-MLLM]
\acro{ACC}[ACC]{Adaptive Cruise Control}
\acro{AD}[AD]{Autonomous Driving}
\acro{ADS}[ADS]{Autonomous Driving Systems}
\acro{AIDE}[AIDE]{Automatic Data Engine}
\acro{AMF}[AMF]{Access and Mobility Management Function}
\acro{AV}[AV]{Autonomous Vehicle}
\acro{Beit}[Beit]{Bidirectional Encoder Representation From Image Transformers}
\acro{BLIS}[BLIS]{Blind Spot Information System}
\acro{CAN}[CAN]{Controller Area Network}
\acro{CLIP}[CLIP]{Contrastive Language–Image Pretraining}
\acro{CNN}[CNN]{Convolutional Neural Network}
\acro{CoT}[CoT]{Chain-Of-Thought}
\acro{CrossViT}[CrossViT]{Cross-Attention Multi-Scale Vision Transformer}
\acro{DRL}[DRL]{Deep Reinforcement Learning}
\acro{E2E}[E2E]{End-To-End}
\acro{EM-VLM4AD}[EM-VLM4AD]{Efficient, Lightweight Multi-Frame Vision Language Model for Visual Question Answering In Autonomous Driving}
\acro{FFNN}[FFNN]{Feed-Forward Neural Network}
\acro{FM}[FM]{Foundation Model}
\acro{GAI}[GAI]{Generative Artificial Intelligence}
\acro{GPU}[GPU]{Graphical Processing Unit}
\acro{GPS}[GPS]{Global Positioning System}
\acro{HDT}[HDT]{Human Digital Twin}
\acro{HiLM-D}[HiLM-D]{High-Resolution Understanding In MLLMs For Autonomous Driving}
\acro{HMI}[HMI]{Human-Machine Interaction}
\acro{IDS}[IDS]{Instruction Detection System}
\acro{IMU}[IMU]{Inertial Measurement Unit}
\acro{LC-LLM}[LC-LLM]{Lane Change Large Language Model}
\acro{Lidar}[LiDAR]{Light Detection and Ranging}
\acro{LIN}[LIN]{Local Interconnect Protocol}
\acro{LKA}[LKA]{Lane Keeping Assist}
\acro{LLM}[LLM]{Large Language Model}
\acro{LLM4AD}[LLM4AD]{Large-scale Language Models For Autonomous Driving}
\acro{MAE}[MAE]{Mean Absolute Error}
\acro{MLLM}[MLLM]{Multimodal Large Language Model}
\acro{MLM}[MLM]{Masked Language Modeling}
\acro{MLP}[MLP]{Lightweight Multilayer Perception}
\acro{MOST}[MOST]{Media Oriented System Transport}
\acro{MPC}[MPC]{Model Predictive Control}
\acro{MuRAG}[MuRAG]{Multimodal Retrieval-Augmented Generation}
\acro{NDD}[NDD]{Natural Denoising Diffusion}
\acro{NN}[NN]{Neural Network}
\acro{NLP}[NLP]{Natural Language Processing}
\acro{ONCE}[ONCE]{One Million Scenes}
\acro{pFedLVM}[pFedLVM]{Personalized Federated Learning Large Vision Model}
\acro{PPO}[PPO]{Proximal Policy Optimization}
\acro{PRISMA}[PRISMA]{Preferred Reporting Items for Systematic Reviews and Meta-Analyses}
\acro{PVT}[PVT]{Pyramid Vision Transformer}
\acro{RAG}[RAG]{Retrieval-Augmented Generation}
\acro{ReAc}[ReAc]{Reasoning and Acting}
\acro{ResNet}[ResNet]{Residual Network}
\acro{RL}[RL]{Reinforcement Learning}
\acro{RLHF}[RLHF]{Reinforcement Learning with Human Feedback}
\acro{RTK}[RTK]{Real-Time-Kinetic}
\acro{SADM}[SADM]{Spatial-Aware Decision Making}
\acro{SAE}[SAE]{Society of Automotive Engineers}
\acro{SAM}[SAM]{Segment Anything Model}
\acro{SMF}[SMF]{Session Management Function}
\acro{SUMO}[SUMO]{Simulation of Urban Mobility}
\acro{T2T-ViT}[T2T-ViT]{Tokens-To-Token Vision Transformer}
\acro{TPU}[TPU]{Tensor Processing Unit}
\acro{TRS}[TRS]{Traffic Rules Satisfaction}
\acro{UDM}[UDM]{Unified Data Management}
\acro{uniMVM}[uniMVM]{Unified Multi-View Model}
\acro{UNITER}[UNITER]{Universal Image-Text Representation}
\acro{UPF}[UPF]{User Plane Function}
\acro{V2D}[V2D]{Vehicle-To-Device}
\acro{V2G}[V2G]{Vehicle-To-Grid}
\acro{V2I}[V2I]{Vehicle-To-Infrastructure}
\acro{V2N}[V2N]{Vehicle-To-Network}
\acro{V2P}[V2P]{Vehicle-To-Pedestrian}
\acro{V2V}[V2V]{Vehicle-To-Vehicle}
\acro{VLM}[VLM]{Vision Language Model}
\acro{VLM4AD}[VLM4AD]{Simulation Of Urban Mobility}
\acro{ViT}[ViT]{Vision Transformer}
\acro{VQA}[VQA]{Visual Question Answering}
\acro{XLM}[XLM]{Any Language Model}
\acro{ZOD}[ZOD]{Zenseact Open Dataset}
\end{acronym}


\section{Related Works}
Recent surveys and studies have explored the application of XLMs within ADS. These investigations assessed the potential of XLMs to enhance the ADS's predictive and decision-making capabilities.
For instance, a categorization and an analysis of recent works applying foundation models to ADS was provided by Gao \textit{et al.} in \cite{gao2024survey}. They established a taxonomy based on modality and functions in ADS, then discussed techniques to adapt FMs to ADS, including in-context learning, fine-tuning, Reinforcement Learning (RL), and visual instruction tuning. They also highlighted the limitations of FMs, such as hallucination, latency, and efficiency, and finally proposed relevant research directions. 
Similarly, \cite{yang2023llm4drive} surveyed the use of LLMs for Autonomous Driving (LLM4AD). Specifically, the authors explored various applications of LLMs within ADS and detailed prominent methodologies in each category. 
They also presented the most recent datasets pertinent to LLM4AD. 
Alternatively, authors of \cite{zhou2024vision} provided a detailed survey on integrating VLMs within ADS. They classified existing studies based on VLM types and application domains while considering five major aspects: Perception and understanding, navigation and planning, decision-making and control, End-to-End (E2E) autonomous driving, and data generation. Moreover, the survey consolidated emerging vision-language tasks and metrics, analyzed classic and language-enhanced AD datasets, explored potential applications and technological advances, and discussed benefits, challenges, and research gaps for ADS.
In \cite{cui2024survey}, Cui \textit{et al.} provided a comprehensive study of integrating MLLMs into ADS. They focused on MLLM's capability to process multimodal data for AD perception, motion planning, and motion control, and identified future research directions in this area. In the same context, authors in \cite{luo2024delving} provided a taxonomy regarding the use of FMs in autonomous vehicles. Moreover, they examined data augmentation and model optimization of LLMs and VLMs to enhance autonomous driving decisions. Finally, authors of \cite{huang2023applications} discussed the architectures of several solutions that apply FMs in ADS.

Although these surveys are informative about the state-of-the-art in XLM and ADS and their integration, several omitted discussing the practicality of using XLMs in a real ADS framework. Moreover, an accurate comparison between the proposed XLM approaches in ADS has not been addressed. Also, in most surveys, only one type of XLM is investigated, e.g., only LLMs or only VLMs. Consequently, we provide here an in-depth examination of the XLM-ADS integration and present a comparative analysis of the XLM-based methods for AD. Our study is more holistic than previous surveys, where we gather and analyze studies covering a larger range of FMs, including LLMs, VLMs, and MLLMs. Furthermore, we target here the most recent state-of-the-art works published within the last two years (2023 and 2024). In Table~\ref{relatedworkspapers}, we highlight the contributions of the previous surveys, compared to ours. 

\begin{table*}
\centering
\caption{Summary of Relevant Survey Papers} 
\label{relatedworkspapers}
\begin{tabular}{|p{0.85cm}|c|p{3.7cm}|c|c|c|c|c|c|c|c|c|}
\hline
\centering \textbf{Ref.} & {\textbf{Year}} & \textbf{Objective} & \textbf{LLMs} & \textbf{VLMs} & \textbf{MLLMs} & \makecell[l]{\textbf{XLM} \\ \textbf{Taxonomy}} & \makecell[l]{\textbf{XLM} \\ \textbf{Archi.}}  & \makecell[l]{\textbf{XLM} \\ \textbf{Datasets}} &  \makecell[l]{\textbf{XLM} \\ \textbf{Simulators} }&   \makecell[l]{\textbf{Future} \\ \textbf{Directions}} \\ \hline

 \centering {\cite{yang2023llm4drive}} & {2023} & {Review of technical achievements of LLMs for AD.} & $\checkmark$ & $\times$  & $\times$ & $\checkmark$ & $\times$  &     $\checkmark$ & $\checkmark$ & $\times$  \\ \hline
\centering \cite{huang2023applications} & 2023 & Architectures of FMs for use in AD.  & $\checkmark$ & $\checkmark$ &    $\checkmark$  & $\times$  & $\checkmark$ & $\times$  & $\times$ & $\times$  \\ \hline 
\centering \cite{gao2024survey} & 2024 & Categorization and analysis of recent works on applying FMs to ADS. &  $\checkmark$ & $\checkmark$ & $\checkmark$  &  $\times$   &  $\times$   &  $\times$  & $\times$  & $\checkmark$ \\ \hline

 \centering \cite{zhou2024vision} & 2024 & Analysis of the potential of VLMs for ADS.  & $\times$ & $\checkmark$ &  
 $\times$ & $\checkmark$  & $\times$ & $\checkmark$    &  $\times$  & $\checkmark$ \\ \hline

\centering  \cite{cui2024survey} & 2024 & Comprehensive study on integrating MLLMs with ADS. & $\times$ & $\times$   &   $\checkmark$   & $\times$ & $\times$  &     $\checkmark$ & $\checkmark$ & $\times$   \\ \hline
 \centering \cite{luo2024delving} & 2024 & Taxonomy of FMs in AD and related optimization methods for enhanced AD decisions.  &   $\times$ & $\times$   &   $\checkmark$   & $\checkmark$  & $\times$  & $\times$  & $\times$  & $\checkmark$\\ \hline 
    
\centering \textbf{Our} \centering \textbf{Survey} & 2024 & An in-depth review of XLMs for ADS.  &  $\checkmark$ & $\checkmark$ &    $\checkmark$ & $\checkmark$ &   $\checkmark$ & $\checkmark$ & $\checkmark$ & $\checkmark$

\\ \hline
\end{tabular}
\end{table*}
\section{Research Methodology}
This section presents our research methodology to gather the relevant papers related to our survey topic and respond to the assessed research questions. 
\subsection{Research Questions}
In conducting this study, our objective is to seek answers to the following questions:
\begin{itemize}
    \item \textbf{RQ0:} What are the fundamental principles of ADS and XLMs? 
    \item \textbf{RQ1:} What are the main AD challenges and issues that can be tackled with XLM-based methods? 
    \item \textbf{RQ2:} How can LLMs be integrated into ADS to improve decision-making, situational awareness, and advanced driver-assistance systems? 
    \item \textbf{RQ3:} How can VLMs be optimized for real-time object detection and identification, and obstacle recognition within AD environments and varying weather conditions? 
    \item \textbf{RQ4:} How can MLLMs be employed to improve visual and linguistic information integration in ADS, and what are the potential applications of MLLMs to enhance human-vehicle interaction systems? 
    \item \textbf{RQ5:} What are the best practices for datasets and tools to evaluate XLMs in the context of ADS?
    \item \textbf{RQ6:} What are the open issues and future research directions for integrating XLMs into ADS?
\end{itemize}

\subsection{Methodology}
For our research methodology, we employed the Preferred Reporting Items for Systematic Reviews and Meta-Analyses (PRISMA) framework \cite{kim2023completeness} to ensure a rigorous and transparent literature review process. PRISMA provides a comprehensive and systematic approach to identify, screen, and include relevant studies to our research questions.
\subsubsection{Identification of Relevant Studies}
The search criteria were applied to identify papers closely aligned with the scope of our study. Accordingly, we conducted a literature assessment using relevant keywords and terms pertinent to our field of interest, i.e., focusing on the deployment of large-scale models in advancing ADS. The search terms and phrases used are, but not limited to, as follows:
\begin{itemize}
    \item ``autonomous driving" and ``large language models" or LLM
     \item ``autonomous driving" and ``vision language models" 
      \item ``autonomous driving" and ``multimodal large language models"
       \item ``autonomous driving challenges'' and ``large scale models" 
       \item ``datasets'', ``LLM'', and ``autonomous driving" 
       \item ``datasets'', ``VLM'', and ``autonomous driving" 
       \item ``datasets'', ``MLLM'', and ``autonomous driving" 
       \item ``Frameworks'', ``LLM'', and ``autonomous driving"
       \item ``Frameworks'', ``VLM'', and ``autonomous driving"
        \item ``Frameworks'', ``MLLM'', and ``autonomous driving"
         \item ``MLLM'', ``autonomous driving", and ``open issues"
\end{itemize}

\subsubsection{PRISMA Process}
\begin{figure}
    \centering
    \includegraphics[trim={1cm 1cm 1cm .5cm},clip,width = 0.99\columnwidth] {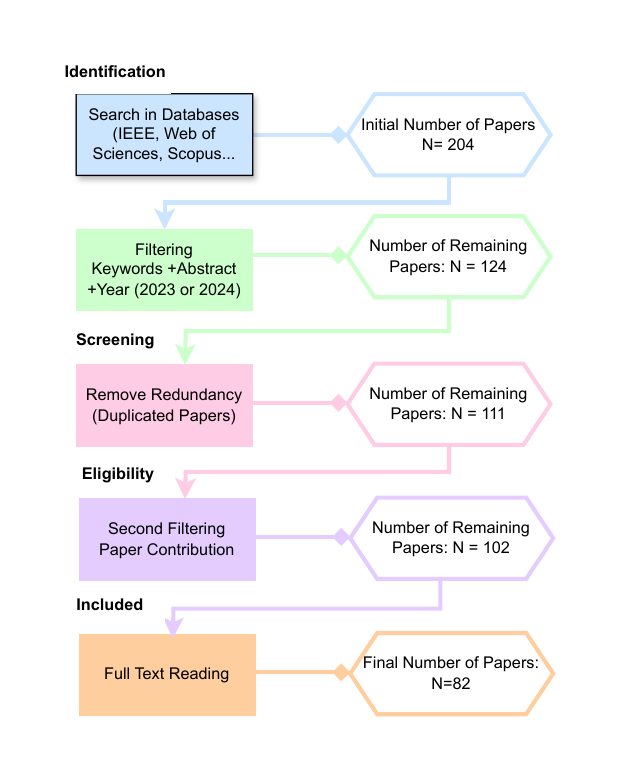}
    \caption{Followed PRISMA steps.}
    \label{Prisma}
\end{figure}

 Accordingly, we started with a broad literature search across multiple databases, followed by the removal of duplicates. Subsequently, we screened the titles and abstracts against predefined inclusion criteria. Full-text papers that met the criteria were then assessed for eligibility. The PRISMA flow diagram, shown in Fig.~\ref{Prisma}, was utilized to document each stage of the review process, providing a clear view from initial identification to final inclusion of studies.
\subsubsection{Selected Papers}
The paper selection process began with an initial search in various databases, yielding $N=204$ papers. Then, a filtering process, based on keywords, abstracts, and publication years (2023 or 2024), reduced the number of papers to $N=124$. To ensure uniqueness, redundant papers were removed, resulting in $N=111$ papers. A second filtering stage was conducted to evaluate the contributions of the papers, further narrowing the list to $N=102$. Finally, a thorough full-text reading was performed and $N=82$ papers were selected for inclusion in the study. 
\section{Background (RQ0)}
\subsection{Brief Overview of ADS}
ADS is considered a significant advancement in automotive technology that revolutionizes the transportation experience \cite{janai2020computer,parekh2022review,ahangar2021survey}. The main factors motivating ADS's development are safety and mobility enhancements, traffic congestion reduction, and fuel efficiency improvement. AD systems are categorized by the Society Of Automotive Engineers (SAE) into six levels of automation, ranging from driver assistance to full self-driving, as follows:
\begin{itemize}
    \item \textbf{Level 0 (No Automation)}:  All driving aspects are the responsibility of the human driver. The vehicle may have systems that provide momentary assistance via certain warnings (e.g., blind spot information system, -BLIS).
    \item \textbf{Level 1 (Driver Assistance)}: The vehicle can assist with either automatic steering or acceleration/deceleration, but the driver always monitors the driving environment.
    \item \textbf{Level 2 (Partial Automation}: The vehicle can control both steering and acceleration/deceleration under certain conditions (e.g., on a highway with clear weather). The driver should remain engaged during the driving process and ready to take over at any moment.
    \item \textbf{Level 3 (Conditional Automation)}: The vehicle can take all driving actions under specific conditions. However, the driver can engage when the system needs or requests it.
    \item \textbf{Level 4 (High Automation)}: The vehicle can perform all driving tasks without human attention or intervention in certain environments or conditions. Nevertheless, human intervention might be required in complex and critical situations.
    \item \textbf{Level 5 (Full Automation)}: In this level, there is no need for a human driver. Indeed, the vehicle can perform and successfully achieve all driving tasks under all conditions.
\end{itemize}


\subsubsection{Architecture of ADS}
\begin{figure*}
    \centering
    \includegraphics[width = 0.93 \textwidth] {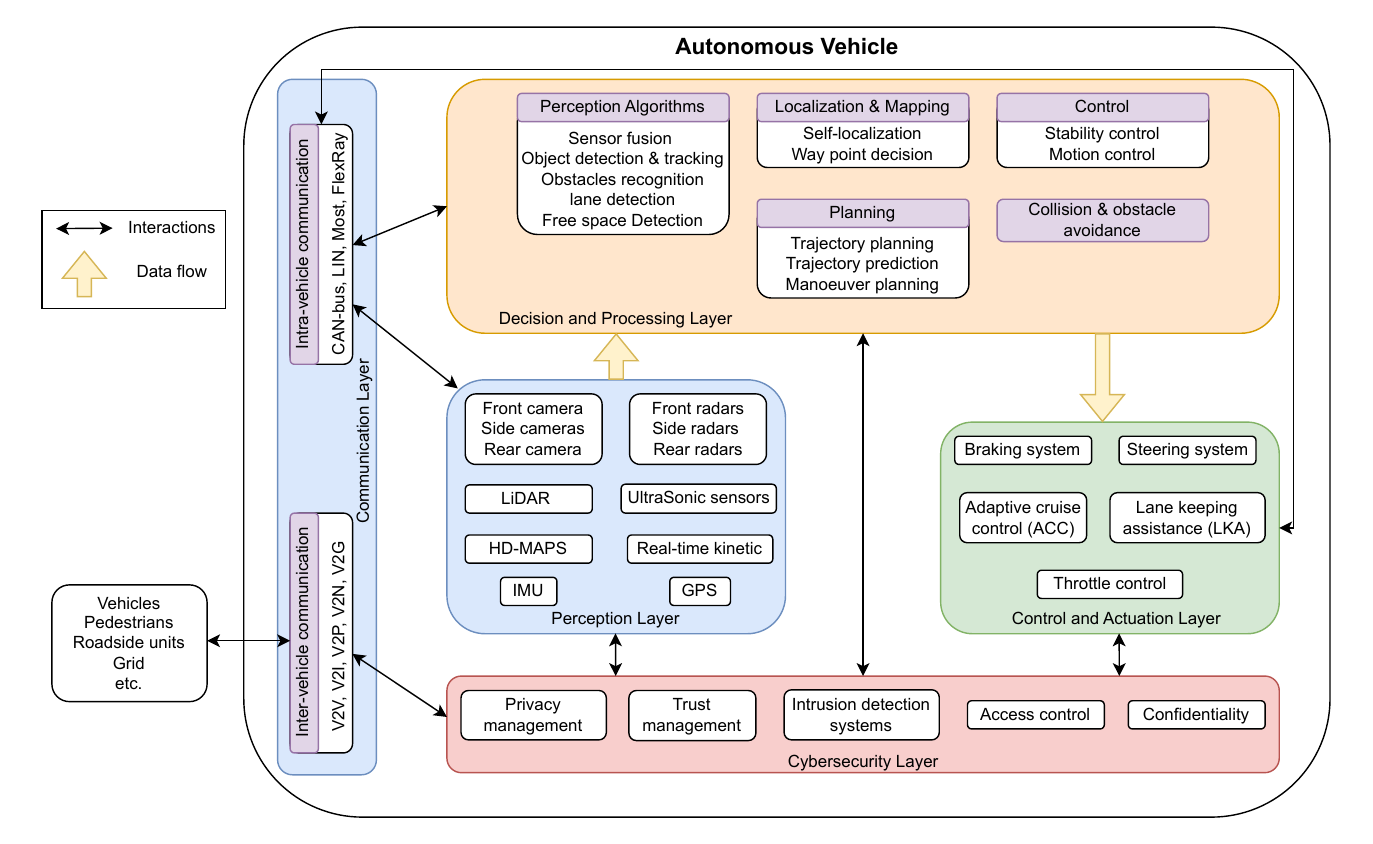}
    \caption{A typical ADS architecture.} 
    \label{ADarchitecture}
\end{figure*}

A typical architecture of ADS should include a combination of sensors, cameras, communication modules, and sophisticated algorithms to navigate and control the vehicle without human intervention. As illustrated in Fig.~\ref{ADarchitecture}, it involves five layers, namely the perception layer, the processing and decision layer, the control and actuation layer, the cybersecurity layer, and the communication layer, described as follows:  
\begin{itemize}
    \item \textbf{Perception Layer:} It involves sensing the surroundings of the Autonomous Vehicle (AV) and detecting the location of the vehicle (self-localization) via various sensors such as front/rear/sides cameras, 360-degree camera, front/rear/sides radars, Light Detection And Ranging (LiDAR), Real-Time Kinetic (RTK), ultrasonic sensors, Global Positioning System (GPS), and Inertial Measurement Unit (IMU). Gathered information from these sensors is forwarded to the processing and decision layer.
    
    
    \item \textbf{Processing and Decision Layer:} It uses gathered data to plan and control the motion and behavior of the AV. This layer is the AV's brain that makes decisions. It integrates several modules and algorithms such as perception algorithms for object detection, object tracking, obstacle recognition, and lane detection, path planning algorithms for trajectory and maneuvers planning, and localization and mapping algorithms for self-localization and way-point decision planning. Decisions are made based on the processing of current and past information including real-time map information, sensed data, traffic details and patterns, and user-provided information.
    
    \item  \textbf{Control and Actuation Layer:} 
It receives information from the processing and decision layer and performs actions associated with the physical control of the AV such as steering, accelerating, and braking. Moreover, its Adaptive Cruise Control (ACC) automatically adjusts the AV's speed to ensure a safe distance from vehicles ahead while the Lane-Keeping Assist (LKA) module ensures that the vehicle remains within its lane.
    \item \textbf{Communication Layer:} It consists of two sublayers: Intra-vehicular communication and inter-vehicular communication. Intra-vehicular communication facilitates interactions between various modules within the AV. Data collected from the perception layer is transmitted to the decision and processing layer, and decisions from this layer are conveyed to the control and actuation layer via intra-vehicular communication modules, using protocols such as Controller Area Network (CAN) bus, Local Interconnect Protocol (LIN), Media Oriented System Transport (MOST), and Flexray. Inter-vehicular communication connects the AV with other vehicles and roadside infrastructure, encompassing various types of communication including Vehicle-To-Vehicle (V2V), Vehicle-To-Infrastructure (V2I), Vehicle-To-Pedestrian (V2P), Vehicle-to-Network (V2N), Vehicle-to-Device (V2D), and Vehicle-To-Grid (V2G). These communication types enable the AV to interact with other vehicles, infrastructure, pedestrians, networks, devices, and the power grid, ensuring a comprehensive and integrated communication system.
    
    \item \textbf{Cybersecurity Layer:} It ensures the safety and reliability of the AV's operations. It encompasses security measures to protect the AV and its modules from cyber threats. It may involve encryption protocols to encode data before transmission, authentication mechanisms to verify the identity of devices interacting with the vehicle, access control modules, and Intrusion Detection Systems (IDS) that monitor the AV's network for signs of malicious activities. The cybersecurity layer interacts with all the other layers to ensure the security of the ADS.
\end{itemize}
\subsubsection{Autonomous driving datasets}
The last decade has been marked by the appearance of several datasets for self-driving applications. Selecting the adequate dataset for a specific AD application needs to consider several parameters, including dataset dimension, environmental diversity, geographical conditions, scenario, city, weather conditions, time of the day, autonomous driving-supported tasks (e.g., perception, planning, or control), and with different street views \cite{liu2024survey}. In the following, we highlight and compare the most relevant datasets.

\begin{itemize}
    \item \textbf{CamVid:} It is the first small dataset with semantically labeled videos, called Cambridge-Driving Labeled Video Database (CamVid) \cite{brostow2008segmentation}. It comprises 701 manually annotated photos with 32 semantic classes.
    
    \item \textbf{KITTI:} The KITTI vision benchmark suite supports multiple computer vision tasks including stereo, optical flow, and 2D/3D object detection and tracking \cite{kitti}. Mainly it focuses on object detection, where 7,481 training and 7,518 testing images are annotated using 2D and 3D bounding boxes. Each image contains up to 30 pedestrians and 15 cars. However, only a few images have pixel-level annotations, thus making KITTI a weak benchmark for semantic segmentation.
    
    \item \textbf{ApolloScape Dataset:} ApolloScape focuses on all aspects of AD, i.e., perception, navigation, and control \cite{xinyuhuang2020apolloscape,apolloscape}. It provides open access to semantically annotated pixel-level street-view images from four regions in China, which are regularly updated.
    
    \item \textbf{NuScenes:} The NuScenes dataset is a large-scale AV dataset \cite{caesar2020NuScenes}. It supports developing and benchmarking ADS, mainly for object detection, tracking, and behavior modeling in complex urban environments and weather conditions. NuScenes is the first multimodal dataset (data gathered with LiDAR, radar, 360-camera, IMU, GPS, and CAN-Bus) to consider geographic diversity since data has been collected from both Boston, USA, and Singapore. It corresponds to 1,000 driving scenes, each is 20 seconds long and provides detailed 3D bounding boxes for 23 object classes, including vehicles, pedestrians, bicycles, traffic cones, and barriers, and high-definition maps including lane markings, road boundaries, crosswalks, and other road infrastructure.

    \item \textbf{Waymo Open:} Compared to other datasets, Waymo offers a large volume of multimodal sensory data collected from multiple locations in diverse weather, geographic, and lighting conditions with complete annotations, including 3D bounding boxes for over 12 million labeled objects, detailed maps, and temporal tracking of data for dynamic objects \cite{mei2022waymo}.
    
    \item \textbf{Argoverse:} It is designed to support AV research in the areas of motion forecasting and 3D tracking through the collection and annotation of high-definition maps, and LiDAR and cameras' data. \cite{chang2019argoverse}. The dataset contains over 320 sequences for tracking and 290,000 scenarios for forecasting, collected from Miami, Florida, Pittsburgh, and Pennsylvania.


    \item \textbf{BDD100K:}  BDD100K is among the largest driving high-resolution video datasets with comprehensive annotations, which includes 100,000 videos of 40 seconds in length each, and 10 tasks dedicated to the evaluation of image recognition algorithms on ADS \cite{yu2020bdd100k}. The dataset provides weather, geographic, and scene diversity as they are collected from more than 50,000 rides covering San Francisco Bay Area, New York, and other regions, in different weather conditions and daytime. 
    This dataset provides rich annotations including 2D bounding boxes for objects, pixel-level annotations for 10 object classes, instance-level segmentation masks for individual objects, and detailed annotations for lane markings.
    \item \textbf{CODA:} This dataset consists of 1,500 real-world
driving scenes \cite{li2022coda}, spanning more than 30 object categories, which are selected from three large-scale AD datasets, namely KITTI, NuScenes, and ONCE \cite{mao2106one}. Each selected scene contains at least one object that is hazardous to AVs or their surroundings. 
 \item \textbf{Zenseact Open Dataset (ZOD):}
ZOD is a multimodal, diverse, and large-scale dataset \cite{alibeigi2023zenseact}, collected over two years and across several European countries. The dataset provides 55.6 hours of annotated data with rich annotations for 2D/3D object detection, traffic sign recognition, road instance and semantic segmentation, and road classification. The dataset is divided into three subsets, including (i) 100,000 independent frames ideal to train and to test perception models on individual images, (ii) 1,473 video sequences, of 20 seconds length each, providing temporal consistency and scene understanding over time, and (iii) 29 driving scenes, with a few minutes length each, ideal to study end-to-end driving tasks.

\item \textbf{Boreas:} It is designed for AD research focusing on weather and seasonal conditions \cite{burnett2023boreas}. Indeed, it has data collected for the same itinerary for over a one-year duration. 
\item \textbf{One Million Scenes (ONCE):} It is a dataset dedicated to 3D object detection in the context of AD \cite{mao2106one}. It includes 7 million images and 1 million LIDAR scenes (larger than NuScenes and Waymo) gathered via a variety of locations, at different times of the day, and in diverse weather conditions. 
\end{itemize}

Table~\ref{comparedataset} summarizes the characteristics of these datasets and highlights their key aspects. 

\begin{table*}[!ht] 
\caption{Summary of Existing AD Datasets}
\centering
\label{comparedataset}
\begin{tabular}{|c|c|c|l|l|l|}
\hline
{\textbf{Ref.}} & {\textbf{Name}} & {\textbf{Data Type}} & {\textbf{Size}} & {\textbf{Features}} & {\textbf{Annotation}} \\
\hline
{\cite{brostow2008segmentation}} & \makecell{CamVid}  & \makecell{Video sequences} & {701 images} & \makecell[l]{960 $\times$ 720 pixels \\per image; Daytime} & \makecell[l]{Pixel-wise\\ semantic segmentation;\\ 32 classes} \\
\hline
{\cite{kitti}} & \makecell{KITTI}  & \makecell{Camera images, LiDAR, GPS,\\ IMU, and  RTK data} & 15,000 images & \makecell[l]{Diverse driving\\ scenarios; Daytime} & \makecell[l]{80,000 3D obj. box.\\ and 15,000 2D obj. box.}  \\
\hline
\makecell{\cite{xinyuhuang2020apolloscape},\\ \cite{apolloscape}} & \makecell{ApolloScape} & \makecell{Camera images, LiDAR,\\ GPS, and IMU data} & 100,000 frames & \makecell[l]{Dense traffic; Varied\\ urban environments \\and weather conditions;\\ Centimeter-level\\ accuracy} & \makecell[l]{Full 3D\\ annotations} \\
\hline
\cite{caesar2020NuScenes} & {NuScenes}  & \makecell{360-camera images, LiDAR,\\ Radar, GPS, and IMU data} & \makecell[l]{1,000 scenes;\\ 23 object classes;\\ 11 map layers} & \makecell[l]{Rich sensor suite;\\ Diverse weather\\ conditions} & \makecell[l]{High-density \\annotations} \\
\hline
\cite{mei2022waymo} & Waymo Open & \makecell{Camera images \\and LiDAR data} & \makecell[l]{1,000 segments;\\ 2,860 seq.;\\ 100,000 images} & \makecell[l]{High-resolution images;\\ Diverse scenes;\\ 8 tracking classes; \\25 semantic classes}  & \makecell[l]{Detailed \\annotations} \\ 
\hline 
\cite{chang2019argoverse} & Argoverse & \makecell{Camera images, LiDAR,\\ GPS, and IMU data} & 324 seq. & \makecell[l]{Motion forecasting;\\ 3D object tracking;\\ Map data} & \makecell[l]{3D bounding box.\\ for several obj.} \\
\hline 
\cite{yu2020bdd100k} & BDD100K  & \makecell{Camera images and GPS data} & 100,000 video seq. & \makecell[l]{Diverse weather\\ conditions; Daytime;\\ Geographic locations} & \makecell[l]{2D bounding boxes;\\ Pixel-level \\annotations \\for 10 object  classes} \\
\hline 
\cite{li2022coda} & CODA  &  \makecell{Camera images and LiDAR data.\\ Raw sensory data from cameras\\ and LIDAR sensors} & 1500 scenes & \makecell[l]{Real-world scenarios \\focused on corner cases;\\ Data integrated from\\ KITTI, NuScenes, \\and ONCE;\\
Contains 34 object classes}
& \makecell[l]{Object-level\\ annotations with\\ bounding box.;\\
Manually verified\\ and corrected\\ annotations} \\
\hline 
\cite{alibeigi2023zenseact} & ZOD & \makecell{Camera images, LiDAR,\\ GPS, and IMU data} & \makecell[l]{100,000 frames;\\ 1,473 seq.;\\ 29 driving scenes} & \makecell[l]{Diverse data over \\time and space;\\ High-resolution sensors} & \makecell[l]{2D/3D bounding box.;\\ Semantic and instance\\ segmentation;\\ 156 classes for\\ traffic signs;\\ Annotated road\\ conditions} 
\\
\hline 
 \cite{burnett2023boreas} & Boreas & \makecell{Camera images, LiDAR,\\ radar, GPS, and IMU data} & \makecell[l]{Data for 350 km\\ of driving} &
\makecell[l]{High-quality data of \\multimodal sensors;\\ Centimeter-level accuracy;\\ Sunny weather} &  \makecell[l]{Ground truth poses;\\ 3D obj. labels}\\
\hline 
\cite{mao2106one} & ONCE  & Camera images, and LiDAR data & \makecell[l]{Scenes 1M, 144 hours,\\ Area 210 km$^2$, \\Images 7M, 417k \\3D boxes, 5 classes} & \makecell[l]{Large scale; High\\temporal and spatial\\ resolution; \\Diverse conditions;\\ Rich annotations}   & \makecell[l]{16,000 scenes \\annotated with \\3D ground truth box.;\\
417k 3D box. \\and 769k 2D box.} 
 \\
\hline
\end{tabular}
\end{table*}
\subsection{Basic Concepts of LLM}
LLMs are advanced language models with massive parameter sizes and exceptional learning capabilities \cite{chang2024survey,zhao2023survey}. 
They can understand contexts and nuances to support tasks across domains such as NLP, which covers text generation, translation, personalized chat-bots, text classification, sentiment analysis, and question answering. In particular, question-answering is a crucial technology for Human-Machine Interaction (HMI), which has proven its usefulness in search engines, intelligent customer service, and question-answering (Q\&A) systems. 
The LLM process starts with pre-training, in which it is exposed to a large dataset of text from a variety of sources, including books, papers, and websites. Unsupervised learning allows the LLM model to anticipate the next word in a phrase based on the context of previous words while comprehensively respecting grammar, syntax, and semantic rules. 

\begin{figure}[t]
    \centering
    \includegraphics[trim={0.5cm 0.5cm 0.2cm 0.5cm},clip,width = 0.99 \columnwidth] {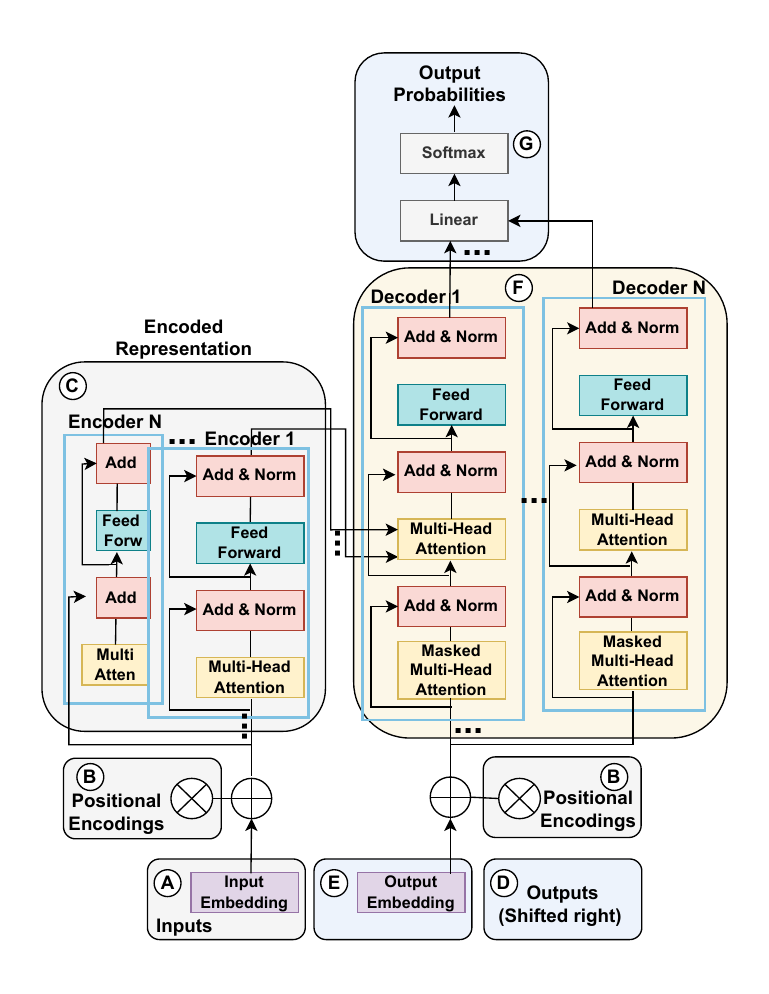}
    \caption{Architecture of Transformer model \cite{raiaan2024review}.} 
    \label{transformer}
\end{figure}
\subsubsection{Basic LLM Modules}
LLMs are built upon several core modules that work together to understand contexts, manipulate, and generate human-like text. The basic LLM modules consist of: 
\begin{itemize}
    \item \textbf{Tokenization:} It parses raw text into tokens or smaller units, such as words, or characters, to be processed by the model.
    \item \textbf{Embedding Layer:} It converts tokens into vectors of a fixed size that capture semantic information. Embedding techniques include word embedding, e.g., Word2Vec, and contextual embedding, e.g., BERT \footnote{BERT stands for Bidirectional Encoder Representations from Transformers. It's a type of LM architecture introduced in October 2018 by researchers at Google.}.
    \item \textbf{Positional Encoding:} It provides the position of tokens in a sequence to understand the sentence structure. Typical encoding techniques include learned positional and fixed positional encodings using sine and cosine functions. 
    \item \textbf{Attention Mechanism:} It allows the model to focus on different parts of the input sequence when generating an output. It analyzes the relationship between tokens in a sequence to facilitate capturing long-range dependencies. Attention mechanisms include self-attention and multi-head attention.
    \item \textbf{Transformer Layer:} It comprises multiple attention and Feed-Forward Neural Network (FFNN) layers to create a deep representation of input data. It could be integrated within the encoder, e.g., BERT, or the decoder, e.g., GPT, or bot, e.g., T5.
     \item \textbf{Normalization Layer:} It accelerates and stabilizes the training process by normalizing inputs across the data mini-batch. 
      \item \textbf{Dropout Layer:} It prevents over-fitting by randomly setting a fraction of input units to zero during training.
       \item \textbf{Output Layer:} It generates the final output from the model, often converting hidden states back into token probabilities. It corresponds to a linear layer followed by a Softmax function \footnote{Softmax function, a.k.a., softargmax or normalized exponential function, converts a vector of numbers into a probability distribution.} for classification tasks.
       \item \textbf{Loss Function:} It calculates the difference between the predicted output and the actual target.
\end{itemize}


\subsubsection{Common LLM Architectures}
The LLM architecture is built around the Transformer framework  \cite{vaswani2017attention}. The two primary parts of a Transformer are the encoder and/or decoder. To find complex correlations between tokens, it breaks down at first the input data into tokens. Then, the latter are subjected to simultaneous mathematical operations. Through this approach, the system is enabled to identify and extract patterns in a way that is comparable to human cognition when faced with a similar question. The Architecture of the Transformer model is illustrated in Fig.~\ref{transformer}. 

Besides, LLMs are characterized by the tasks they are designed for, including text generation, text classification, and text summarizing, and their training strategies, e.g., self-supervised, unsupervised, distillation, fine-tuning, etc. LLMs deploy large-scale text corpora during the training process that necessitates significant computational resources of advanced hardware like Tensor Processing Units (TPUs) and Graphical Processing Units (GPUs). Therefore, the required hardware is another criterion that needs to be considered when selecting LLMs for specific tasks.

\begin{table*}[t]
\centering
\caption{Summary of LLM architectures}
\label{tablellm}
\begin{tabular}{|p{0.5cm}|c|p{1.3cm}|p{1.5cm}|p{2cm}|p{2.9cm}|p{1.8cm}|p{2.2cm}|} 
\hline
\textbf{Year} & \textbf{Model Name} & \textbf{Model Archi.}  & \textbf{Model Parameters} & \textbf{Pre-training Method} & \textbf{Pre-training Datasets} & \textbf{Hardware Specs.} & \textbf{Training Duration} \\ \hline
2018 & BERT & Encoder   & 110-340 Million & Self-supervised & BookCorpus and English Wikipedia & Nvidia A100 and V100 & Variable (Depends on parameter scale) \\ \hline
2018 & GPT-1 & Decoder  & 117 Million & Self-supervised & BookCorpus and English Wikipedia & - & - \\ \hline
2019 & DistilBERT & Encoder & 66 Million & Self-supervised; Distillation  & BookCorpus and English Wikipedia &  Nvidia V100 & 90 hours \\ \hline
2019 & RoBERTa & Encoder & 125-355 Million & Self-supervised & BookCorpus, Openwebtext, CC-news, and stories & 6144 TPUs v4 & $\approx$ 2 weeks \\ \hline
2019 & Sentence-BERT & Encoder & 110 Million & By fine-tuning & SNLI and Multi-Genre NLI &  Nvidia V100& - \\ \hline
2019 & BART & Encoder-Decoder  & 140-400 Million & Self-supervised & Books corpus, Openwebtext, CC-news and stories & - & - \\ \hline
2019 & T5 & Encoder-Decoder  & 60 Million - 11 Billion & Self-supervised & Colossal Clean Crawled Corpus & 1024 TPUs v3 & - \\ \hline
2019 & GPT-2 & Decoder  & 1.5 Billion & Self-supervised & WebText & - & - \\ \hline
2020 & GPT-3 & Decoder & 175 Billions & Unsupervised & Common Crawl, WebText2, Books1/2, and Wikipedia & Nvidia A100 & - \\ \hline
2020 & wav2vec2 & Encoder-Decoder & 227-896 Million & Self-supervised & LibriSpeech and Unlabeled Audio Data & - & - \\ \hline
2021 & GLM & Encoder & 110 Million - 130 Billion & Self-supervised & BookCorpus and English Wikipedia & 1024 TPUs v4 & 60 days \\ \hline
2021 & HuBERT & Encoder-Decoder& 281 Million - 2.8 Billion & Self-supervised & Libri-Light and LibriSpeech & - & - \\ \hline
2022 & InstructGPT & Decoder & 175 Billion & Unsupervised RLHF & Common Crawl, WebText2 Books1/2, and Wikipedia & 992 Nvidia 80G A100 & - \\ \hline
2022 & PaLM & Decoder & 54 Billion & Unsupervised & 780 billion tokens from social media, webpages, books, Github, multilingual Wikipedia, and news & 6144 TPUs v4 & 120 days \\ \hline
2023 & Whisper & Encoder-Decoder & 39-1150 Million & Self-supervised & - & - & - \\ \hline
2023 & LLaMA & Decoder & 7-70 Billion & Self-supervised & Common Crawl, C4, Github, Wikipedia, Books, ArXiv, and StackExchange & 2000 Nvidia 80G A100 & 21-25 days \\ \hline
\end{tabular}
\end{table*}

\subsubsection{Interacting with LLMs}
It can be achieved using techniques like prompt engineering, fine-tuning, zero-shot, and Reinforcement Learning From Human Feedback (RLHF). These methods enhance the model’s performance and adapt it to specific tasks or domains. They are described as follows:  
\begin{itemize}
    \item \textbf{{Prompt Engineering}:}
Prompt engineering for LLMs involves crafting inputs (prompts) to guide the model in generating the desired responses \cite{marvin2023prompt}. Several strategies and best practices have been developed for effective prompt engineering, including: 
\begin{itemize}
    \item \textit{Chain-of-Thought (CoT) prompting:} Using a sequence of
interconnected prompts to guide coherently and logically the language model’s responses.
    \item \textit{Self-consistency:} Using prompts to encourage the language model to generate responses that are consistent with its previous ones.
 \item \textit{Knowledge generation prompting:} Using prompts to encourage the
language model to generate new insights based on existing knowledge and understanding.
 \item \textit{Reasoning and Acting (ReAc):} Using prompts to encourage the language model to reason about a given situation and generate appropriate actions or responses.
 \item \textit{Contextual prompting:} Providing additional context to
the language model to guide it to generate more relevant responses.
 \item \textit{Dynamic prompting:} Dynamically adjusting the prompt based on the language model’s previous responses to improve its performance over time.
 \item \textit{Transfer learning prompting:} Using transfer learning to adapt a pre-trained language model to new tasks using crafted prompts.
\end{itemize}
    \item \textbf{{Fine-tuning}:} Fine-tuning LLMs involves training them on a specific dataset to customize them to respond to requests in a specific context \cite{han2024parameter}. It stimulates the model to generate consistent outputs and reduces hallucinations. Fine-tuning can be realized in an unsupervised, supervised, or instruction-based manner.

    \item \textbf{{Zero-shot, One-shot, and Few-shot Learning}:} Recent studies suggested that LLMs exhibit high levels of generalization, enabling them to apply their acquired knowledge to new tasks not included in their original training process. This capability is known as zero-shot learning \cite{sun2023evaluating}. When the model is provided with a single example to illustrate the task, it corresponds to one-shot learning while few-shot learning is when the model is provided with a few examples to better understand the task requirements and format. 
    
    \item \textbf{{Reinforcement Learning from Human Feedback}:} RLHF is an advanced technique of fine-tuning, where feedback collected from users regarding the model's responses is used to enable the LLM model to learn from it and improve its future responses \cite{chan2024dense}.
   
 \item \textbf{{Multi-modal Integration}:}
 LLMs could be enhanced by integrating them with other data modalities such as images, audio, or structured data. Indeed, combining image data with textual prompts leads to a comprehensive model capable of understanding and generating responses based on both text and visual inputs. Also, integrating structured data, e.g., tables and databases with textual inputs, leads to more informed and accurate responses.
 \end{itemize}
In Table~\ref{tablellm}, we summarize the main characteristics of common LLM architectures.
 
\subsubsection{Workflow for Efficient  LLM Utilization}
A typical workflow for efficient LLM utilization involves the following steps:
\begin{itemize}
\item \textbf{{Task identification}:} Determine and identify the specific task that the LLM should perform.
   \item \textbf{{Model selection}:} Select the right pre-trained LLM that aligns with the task requirements of your application. Several LLMs are available such as GPT-4, BERT, and RoBERTa, where each model has different strengths and weaknesses. 
   \item \textbf{{Model fine-tuning}:} This step corresponds to the customization of the pre-trained model for a specific task by, for instance, training it on a particular dataset. This step leads to hyperparameter adjustments, e.g., learning rate, batch size, and epochs.
    \item \textbf{{Model evaluation}:} This step implies measuring the model's performance using metrics such as accuracy, recall, precision, and F1-score on a test dataset to ensure that it meets the task's requirements.
    \item \textbf{{Model deployment}:} It consists of integrating the fine-tuned model into the targeted application or system, setting up APIs or user interfaces, and establishing monitoring and logging for production performance.
     \item \textbf{{Model improvement}:} It corresponds to the incorporation of user feedback, then updating the model with new data to enhance its future performance.
\end{itemize}

\subsection{Basic Concepts of VLMs}
Vision language models, a.k.a., VLMs, are advanced Neural Networks (NNs) to process and understand visual data such as images and videos. They demonstrated high performances in several computer vision tasks, e.g., object detection, segmentation, and image classification. Like LLMs, VLMs provide discriminative and generative tasks.

\begin{figure}
        \centering
    \includegraphics[trim={0.5cm 0.2cm 0.2cm 0.1cm}, clip,width = 0.99 \columnwidth] {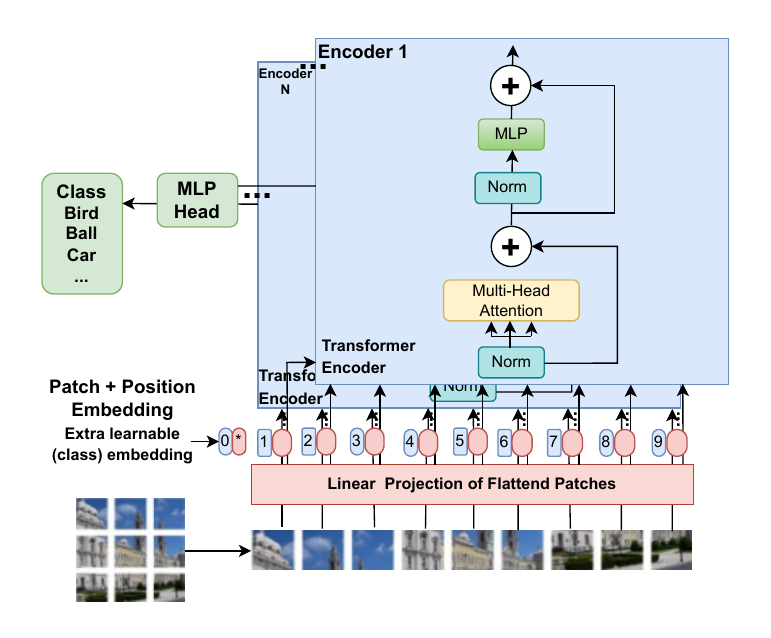}
    \caption{Structure of ViT \cite{dosovitskiy2020image}.}
    \label{VIT}
\end{figure}

The fundamental building blocks of VLMS are Convolutional NNs (CNNs). 
An interesting adopted CNN structure for VLMs is the Residual Network (ResNet). The latter enables the training of deeper NNs than CNNs. Moreover, Transformer models have been used for vision-based tasks. Specifically, Vision Transformers (ViTs) divide an image into fixed-size patches, linearly embed them into vectors, and then process them by Transformer encoders and self-attention mechanisms to capture relationships between patches, as shown in Fig. \ref{VIT}. Positional encodings are added to the patch embeddings to retain spatial information, allowing the model to differentiate between spatial positions. 
 

Following ViT, several Vision Transformer variants have been proposed to enhance training and inference efficiency in image recognition and generation tasks. Among the ViT architecture variants, we highlight the following. First, DeiT incorporated distillation strategies in ViT to outperform standard CNNs without requiring pre-training on large-scale datasets \cite{touvron2021training}. Also, Tokens-To-Token Vision Transformer (T2T-ViT) \cite{yuan2021tokens}, Swin Transformer \cite{liu2021swin}, ViT Masked AutoEncoder (VitMAE) \cite{he2022masked}, VitDet that combines ViT and Deit \cite{li2022exploring}, Pyramid Vision Transformer (PVT) \cite{wang2021pyramid}, and Cross-Attention Multi-Scale Vision Transformer (CrossViT) \cite{chen2021crossvit} focused on improving the network architecture of ViT, mainly to make it lighter and fast-processing. In contrast, AdaViT \cite{meng2022adavit} reduced the computation cost of ViT through E2E adaptive computation, while DETR \cite{carion2020end} achieved it with a hybrid CNN-Transformer architecture.  
In \cite{xie2021segformer}, SegFormer has been proposed, which is a semantic segmentation model that unifies Transformers with Lightweight Multilayer Perception (MLP) decoders. In addition, DINOv2 exploits training on a large ViT model, before distilling it into smaller models that can achieve higher performances \cite{oquab2023dinov2}.
In addition, the bidirectional encoder representation from image Transformers, a.k.a., Beit, introduced a self-supervised vision representation model \cite{bao2021beit}. Specifically, it executes masked image modeling to pre-train vision Transformers and fine-tune the model parameters.  
Finally, the LVM can learn exclusively from visual data \cite{bai2023sequential}. 
In Table~\ref{tablelvm}, we summarize the characteristics of VLM architectures.

\begin{table*}[htbp]
\centering
\caption{Summary of VLM architectures }
\label{tablelvm}
 \begin{tabular}{|p{0.5cm}|c|p{2cm}|p{2.8cm}|p{1.5cm}|p{2.8cm}|p{3.2cm}|}
\hline
\textbf{Year} & \textbf{Model Name} & \textbf{Model Arch.} & \textbf{Oriented Tasks} & \textbf{Model Parameters} & \textbf{Pre-training Method and Dataset}  & \textbf{Testing Datasets} \\ \hline
 2020 & DETR  & Encoder-Decoder & Object detection; Instance segmentation; Panoptic segmentation & 40 Million & Supervised; COCO-2017 & COCO-2017 \\ \hline
 2020 & ViT  & Encoder & Image classification & 86-632 Million & Supervised and self-supervised; ImageNet-21K & ImageNet-1K \\ \hline
 2021 & DeiT & Encoder & Image classification & 5-88 Million & Distilled; ImageNet-1K & ImageNet-1K, CIFAR-10, CIFAR-100, Flowers, Cars, iNat-18, and iNat-19 \\ \hline
 
 2021 & SegFormer & Encoder-Decoder & Segmentation & 3.7-82 Million & Supervised; ImageNet-1K & Cityscapes, ADE20K, and COCO-Stuff \\ \hline
 2021 & Swin Transformer & Encoder & Image classification; Object detection; Semantic segmentation; Video classification & 26 Million & Self-supervised; ImageNet-22K & ImageNet-1K, ADE20K, COCO, and Object 365 v2 \\ \hline
 2021 & BEiT & Encoder & Image classification; Semantic segmentation & 86-632 Million & Self-supervised; ImageNet-1K & ImageNet-1K and ADE20K \\ \hline
2021 & T2T-ViT & Transformer-based & Image classification & 6.9M - 65 Million & Trained from scratch on ImageNet-1k	Tokens-to-Token (T2T) mechanism  & ImageNet-1k\\ \hline  			
2021 & PVT & Encoder & Classification; Object detection; Semantic segmentation  & PVT-S: 25 Million, PVT-M: 44 Million & Supervised & ImageNet-1K, COCO, and ADE20K \\ \hline 		
2021 & CrossViT & Cross-attention vision Transformer & Image classification & Variable (Multi-scale) & Supervised; ImageNet-1K & ImageNet-1K	\\ \hline
2022 & AdaViT & ViT & Image classification & Variable (Adaptive) & Supervised; ImageNet-1K	 & ImageNet-1K \\ \hline
 2022 & ViTMAE & Encoder & Image classification & 86-632 Million & Self-supervised; ImageNet-1K & COCO, ADE20K, and iNat Places \\ \hline
 2022 & ViTDet & Encoder & Object detection & 86-632 Million & MAE minimization; ImageNet-1K & COCO \\ \hline  
 2023 & DINOv2 & Encoder & Image classification & 1.1 Billion & Self-supervised; LVD-142M & ImageNet-1K, ImA, and Oxford-II \\ \hline
2023 & LVM  & Encoder-Decoder & Semantic segmentation; Depth estimation; Surface normal estimation; Edge detection & 300 Million - 3 Billion & Self-supervised; UVD-v1 & Kinetics-700 and ImageNet. \\ \hline

\end{tabular}
\end{table*}




\subsection{Basic Concepts of MLLMs}
MLLMs build upon the foundation of traditional LLMs by enhancing their ability to process and manage information from various sources, including text, images, and videos. This extension allows MLLMs to deliver more contextually relevant and accurate responses. 

\begin{figure*}[t]
        \centering
    \includegraphics[width = 0.97 \textwidth] {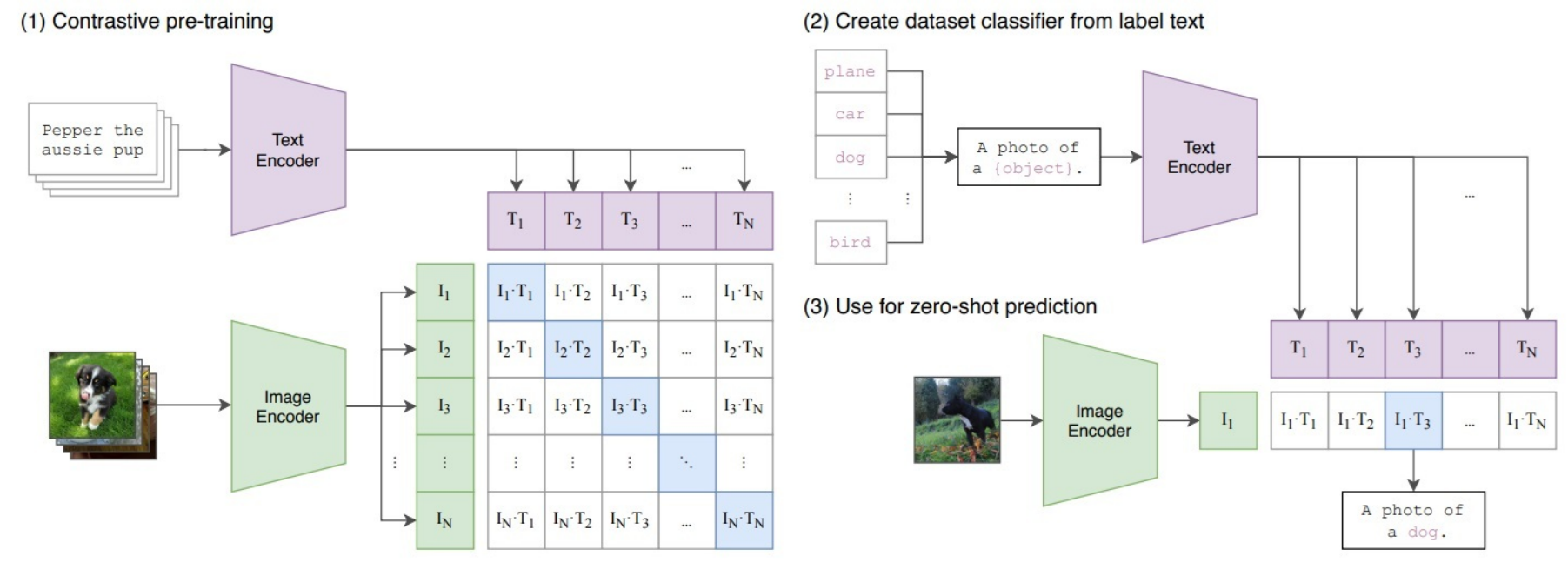}
    \caption{Architecture of CLIP framework \cite{radford2021learning}.} 
    \label{clip}
\end{figure*}

\begin{figure*}[t]
        \centering
    \includegraphics[trim={1cm 0.7cm 1.2cm .5cm},clip,width = 0.9 \textwidth] {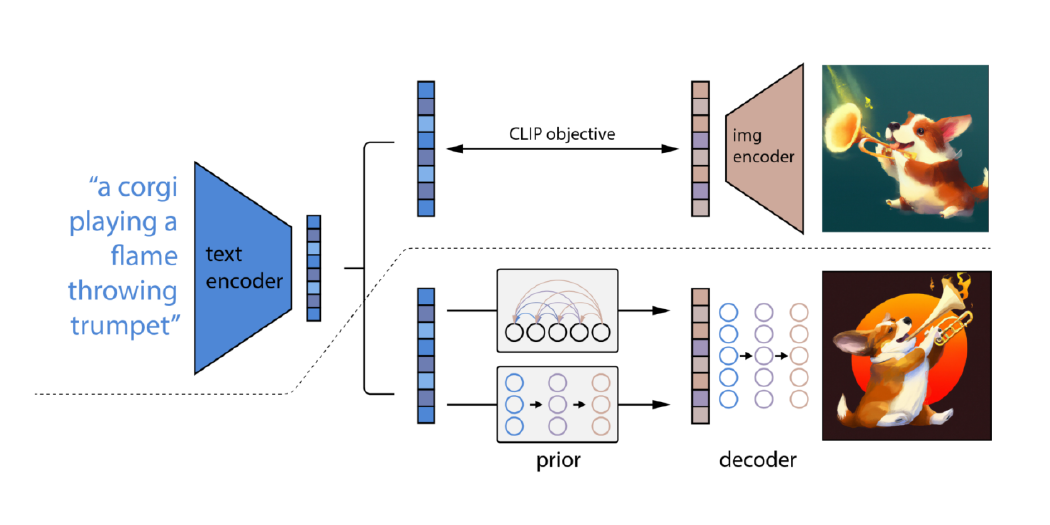}
    \caption{Architecture of DALL-E2 framework \cite{ramesh2022hierarchical}.} 
    \label{DALLE}
\end{figure*}

MLLMs have three key architectural components that enable them to process and integrate information from multiple modalities, namely (i) a pre-trained modality encoder, (ii) a pre-trained LLM model, and (iii) a modality interface that links the previous two components. Recently, contributions to MLLMs have converged to Transformer as the main framework for multi-modal data interaction, including text-to-image and image-to-text retrieval, image captioning, and image/text generation. Accordingly, OpenAI has developed two multi-modal models, called Contrastive Language–Image Pertaining (CLIP) and DALL-E \cite{radford2021learning,ramesh2021zero}. CLIP integrates visuals and their textual information in training and contrastive learning to create a shared embedding space for texts and images. Hence, CLIP can perform zero-shot learning to recognize and classify new images. As shown in Fig. \ref{clip}, CLIP architecture involves three main modules: (1) Constrastive pre-training to learn matching images with their corresponding text description using text and image encoders, (2) classifier creation for a new task without additional training on task-specific data. It requires two steps. First, label text encoding where desired labels for the new task are converted into text descriptions. Then, text encoder utilization where text descriptions are fed through the pre-trained text encoder to generate their respective embeddings. Finally, (3) zero-shot prediction, in which a new image is encoded to generate its embedding, then the latter is compared to the embeddings of the label texts generated in the previous stage, hence predicting the right label for the embedded image.
Moreover, DALL-E generates images from textual descriptions, emphasizing creative visual content synthesis \cite{ramesh2022hierarchical}. It is built with three modules, namely (1) a Transformer model that captures the long-range dependencies in sequences, (2) a text encoder to convert text into a series of embeddings that capture the semantic meaning of the text, and (3) an image encoder that generates images from embeddings. 
{DALL-E2 and DALL-E3 are new versions of DALL-E \cite{ramesh2022hierarchical}. As an example, we illsutrate in Fig.~\ref{DALLE} the DALL-E2 architecture. The latter uses an encoder-decoder pipeline that encodes the text description into a CLIP embedding representing both the text and image content. Then, the model decodes the embedding back to an image using a diffusion model.}
Furthermore, the Universal Image-Text Representation (UNITER) is among the first MLLMs to combine images and text embedding \cite{chen2020uniter}. It has been trained for masked language modeling, image-text matching, and masked region modeling. Similarly, the Simple Visual Language Model (SimVLM) has been proposed for the joint representation of images and text, while reducing the training complexity using large-scale weak supervision \cite{wang2021simvlm}.
In addition, BEIT-3 was proposed as a
general-purpose MLLM, as it uses multiway Transformers for deep fusion and modality-specific encoding \cite{wang2023image}, while PaLM-E is a large embodied multimodal model that addresses a variety of embodied reasoning tasks such as robotics tasks \cite{driess2023palm}. The latter is among the large MLLMs with 562 billion parameters. 
Finally, authors of \cite{huang2024language} proposed KOSMOS-1, a MLLM that can perceive
general modalities and follow instructions, e.g., zero-shot.

To achieve efficient MLLMs, several issues need to be addressed by researchers, including efficient cross-modal interactions, the variability of data quality, and the computational complexity within large-scale models. Efforts to reduce the computational overhead, enhance fusion mechanisms, and increase the robustness of MLLMs, are ongoing, leading until now to several improvements. In Table \ref{mllmarchtable}, we summarize the characteristics of selected MLLM architectures \cite{xu2024survey,wang2024exploring,wu2023multimodal}, while Table \ref{compare} presents a comparison between LLM, VLM, and MLLM models. 



\begin{table*}
\centering
\caption{Summary of MLLM architectures}
\label{mllmarchtable}
\begin{tabular}{{|p{0.4cm}|c|p{1cm}|p{2.6cm}|p{3cm}|p{1.4cm}|p{4.5cm}|}}
\hline
\textbf{Year} & \textbf{Model Name} & \textbf{Model Archi.} & \textbf{Knowledge Representation} & \textbf{Learning Objectives} & \textbf{Model Parameters} & \textbf{Pre-training Datasets} \\
\hline
2020 & UNITER  & Encoder & Complex visual feature embedding & Multiple objectives (e.g., MLM,Image-text contrast(ITC)) & 110 Million & COCO, Visual Genome Conceptual Captions, and SBU Captions   \\
\hline
2021 & SimVLM  & Encoder-Decoder & Pre-trained visual and language models & Generative tasks & \{86, 307, 1200\} Million & ALIGN 
\\
\hline
2021 & CLIP & Encoder & Image-text pairs; Unsupervised techniques & ITC & 400 Million – 1.6 Billion & CLIP-ViT-B:400M, CLIP-RN50:400M, CLIP-RN101:500M, and CLIP-RN50*4:1.6B \\
\hline
2021 & DALL-E 2  & Encoder-Decoder & Diffusion model conditioned on CLIP image embeddings & Generative tasks with text prompts & 12 Billion &  MS-COCO and YFCC100M \\
\hline
2022 & BEiT-3  & Encoder & Shared multiway Transformer; Masked data pre-training & Masked Language Modeling; Masked Visual Modeling & 2 Billion & ImageNet-1K, JFT-3B, COCO, Visual Genome, and Visual Question Answering v2.0  \\
\hline

2023 & PaLM-E  & Encoder-Decoder & Combines PaLM and ViT-22B models & Object detection; Scene classification; Language tasks (e.g., code generation, math equations) & 562 Billion & LAION-400M, WebVid-10M, HD-VILA-100M, and ACAI-1M  \\
\hline

2023 & MiniGPT-4  & Encoder-Decoder  & Aligning a frozen vision encoder with a LLM & Cross-modal generation & 13 Billion & Conceptual Captions, SBU Captions, and LAION\\
\hline
2024 & KOSMOS-1  & Encoder-Decoder & Multimodal integration; In-context learning & Visual dialogue; Image captioning; Zero-shot image classification & 1.6 Billion &  Large-scale Web image-text pairs \\
\hline
\end{tabular}
\end{table*}



\begin{table*}
    \centering
    \caption{Comparison between LLM, VLM, and MLLM}
    \begin{tabular}{|p{2.2cm}|p{4.7cm}|p{4.7cm}|p{4.7cm}|}
        \hline
        \textbf{Aspect/XLM Type} & \textbf{LLM} & \textbf{VLM} & \textbf{MLLM} \\
        \hline
        \textbf{Definition} & Models designed to process and generate text & Models designed to process and interpret visual data, while adding textual interpretation as a secondary step & Models integrating and processing multiple data including text, images, audio, etc   \\
         \hline
        \textbf{Main Functions} & Natural language understanding and generation  &  Vision-related tasks, e.g., image recognition and object detection & Understanding and generating outputs across different modalities, e.g., text, image, and audio \\
        \hline
        \textbf{Input Types}  & Text &  Images, videos, and text (optional) & Images, videos, audios, and text \\
        \hline
        \textbf{Output Types} & Text &  Labels, segmentation maps, and bounding boxes & Images, videos, audios, and text \\
         \hline
        \textbf{Relevant Applications} & Text generation, translation, summarization, question answering & Image classification, object detection, facial recognition, video analysis & Images captioning, video describing, speech-to-text and text-to-speech\\
        \hline
        \textbf{Model Archi.} & Transformer-based architectures &  CNN and ViT-based architectures & LLM and VLM hybrid architectures supported by Transformers \\
         \hline
        \textbf{Training Data} & Large text corpora  &  Large image and/or video datasets & Large multi-modal datasets (e.g., image-text pairs)\\
        \hline
        
        \textbf{Strengths} & Proficient in language understanding and generation & High accuracy in visual tasks and robust feature extraction from images & Can handle complex tasks requiring understanding of multiple data types\\
         \hline
        \textbf{Weaknesses} & Limited to text-based inputs and outputs& Limited mainly to visual-based inputs and outputs & More complex to train and integrate with the highest computational costs \\
        \hline
       
        \textbf{Cross-domain Transfer} & Transfer learning within text domains &  Transfer learning within vision domains & Transfer learning across several text and/or vision domains \\
                \hline
    \end{tabular}
        \label{compare}
\end{table*}

\section{XLMs to Mitigate ADS Issues (RQ1)}
To guarantee the development of safe, reliable, and efficient ADS, and to ensure seamless integration of AVs into transportation networks, several challenges should be addressed. We describe below the most relevant challenges related to ADS. 

\subsection {Multi-Modality of Inputs and Sensors Fusion} 
To perceive their environment, AVs rely on a variety of sensors, including visual sensors (using cameras), proximity sensors (using LiDAR and Radar), ultrasonic sensors, navigation signals (with GPS), language instructions, and HD-Maps. Moreover, for robust and efficient perception, the large volume of gathered multi-modal data should be combined. This task is known as sensor fusion. Indeed, accurate scene understanding requires that multi-modal data is synchronized to the same spatial and temporal coordinates. However, given their heterogeneity, achieving efficient sensor fusion is challenging. 
Recent studies started investigating mechanisms of multi-sensor fusion and cooperative perception. They were surveyed in \cite{xiang2023multi,singh2023transformer,zhao2023potential,hasanujjaman2023sensor}. However, with the advent of MLLMs, further investigation is needed to enable more efficient sensor fusion. For instance, authors of \cite{choi2023semantics} proposed semantics-guided Transformer-based sensor fusion to improve way-point predictions.   


\subsection{Safety and Reliability}
Designing systems that can manage in real-time sensor failures, unexpected situations and weather conditions, and software bugs, without jeopardizing AD safety, is a critical issue. 
To overcome such challenges, ADS algorithms should be trained in a diversity of situations and conditions to improve their environment perception. Accordingly, XLMs are a key enabler for enhanced perception and decision-making, particularly in critical situations. In this context, authors of \cite{nouri2024engineering} proposed a solution that generates safety requirements via the use of a pipeline of prompts and LLMs that receive item definitions. The pipeline also reviews the requirements' dataset to identify redundant or contradictory requirements.  
In \cite{wang2023empowering}, the authors examined how LLMs can be integrated into ADS. They proposed techniques that use LLMs to make intelligent decisions in behavioral planning and included a safety verifier for contextual learning to improve driving performance and safety. 
Finally, authors in \cite{wang2023accidentgpt} presented AccidentGPT, an E2E accident analysis and prevention framework based on the perception of the vehicle-to-everything environment and the use of MLLMs. Their objective was to improve traffic safety during the transition from manual driving to AD.



\subsection {Complex Urban Environments} 
Navigating through complex urban environments with dynamic components such as pedestrians, vehicles, and cyclists, presents numerous challenges for ADS. 
Furthermore, the latter should understand and adhere to local traffic laws and conventions, which can vary between regions.
Accordingly, authors of \cite{luo2024delving} investigated multi-modal multi-task visual understanding FMs, specifically designed for road scenes. These models leverage multi-modal and multi-task learning capabilities to process and fuse data from diverse sources, enabling them to handle various driving-related tasks with adaptability.  


\subsection {Data Privacy and Security} 
Given the large amount of data generated and processed by AVs, there is a serious risk of data/sensor breaches, alteration, and/or eavesdropping. XLMs can be exploited to protect the AVs' data. For instance, authors of \cite{aldeenwip} proposed the use of MLLMs to mitigate natural denoising diffusion (NDD) attacks on traffic signs and integrate them into ADS.


\subsection {Human-Machine Interaction} 
The development of AVs brings new HMI opportunities and challenges. 
Indeed, as AVs evolve, understanding and responding to human intent becomes a significant requirement. Therefore, a smooth and intuitive interaction between AVs and drivers, passengers, and other road users, is required to realize large-scale adoption of ADS. Designing interfaces that allow passengers to understand the AV's actions and intentions is necessary. The integration of chat-bots, voice-to-text, text-to-voice, text-to-image, and image-to-text functionalities into AVs would enhance HMI and make it more intuitive and natural. In this context, authors of \cite{cui2023human} studied how integrating LLMs with Human Digital Twin (HDT) can change HMI for AD. 
Similarly, Yang \textit{et al.} highlighted in \cite{yang2024human} the benefits of integrating LLMs into ADS. 
Specifically, they conducted experiments using various LLM models and prompt designs to evaluate their effectiveness in few-shot multivariate binary classification tasks. 
Results have shown that GPT-4 is the most accurate one in task understanding and responding, compared to other LLMs such as CodeLlama. Finally, the authors of \cite{xu2023drivegpt4} introduced in DriveGPT4, an interpretable E2E ADS based on LLM. They showed that DriveGPT4 can process textual queries and multi-frame video inputs, thus facilitating the interpretation of vehicle actions with reasoning.





\section{Proposed Taxonomy of XLM-based Approaches for Autonomous Driving}
 This section introduces our proposed taxonomy for the application of XLMs in the context of AD. The taxonomy is designed to address three research questions, namely RQ2, RQ3, and RQ4. The detailed approaches related to the proposed taxonomy will be discussed in the remaining sections of the survey. 
 Specifically, section VII divides the use of LLMs in AD into four main categories as follows:
 \begin{enumerate}
     \item \textbf{Prompt engineering-based methods:} Within this category, studied contributions are classified according to the provided AD tasks, which are planning and control, perception, multi-tasking, and question-answering. 
     \item \textbf{Fine-tuning-based methods of pre-trained models:} Under this category, we consider approaches that fine-tuned pre-trained LLMs for precise planning and control tasks, lane change maneuvers, and path planning.
     \item \textbf{RLHF-based methods:} We classified the studied solutions within this category into approaches that utilize RLHF to improve decision-making in planning and control, and to generate scenarios to test and improve the AV's response to critical situations. 
     \item \textbf{LLM and GAI-based methods:} This category explores the integration of LLMs with GAI to design advanced AD solutions.
 \end{enumerate} 
 
Subsequently, Section VIII categorizes the use of VLMs in AD into two categories as follows:
\begin{enumerate}
    \item \textbf{Prompt engineering-based methods:} Here, we discuss solutions to improve the AV's perception of its environment using VLMs including VLMs' implementations to answer visual and scene interpretation queries.
     \item \textbf{Fine-tuning-based methods:}
This category includes frameworks based on VLMs and makes use of fine-tuning techniques to improve accuracy in perceiving the driving environment, as well as frameworks that adjust VLMs for improved question-answering tasks related to visual data.
\end{enumerate}

Finally, Section IX focuses on the integration of MLLMs in ADS. The studied methods in this section are classified into four categories as follows:
\begin{enumerate}
    \item \textbf{Prompt engineering-based methods:}
Under this category, studied approaches include solutions that use MLLMs to enhance the environment perception and those that improve the performances of Q\&A tasks. 
    \item \textbf{Fine-tuning-based methods:} Studied frameworks in this category aim to enhance planning and control tasks, trajectory planning, perception tasks, and question-answering.

    \item \textbf{RLHF-based methods:} Studies addressed here use RLHF to train agents for better perception, planning, and control, and to enhance waypoint prediction.
    \item \textbf{MLLM and GAI-based methods:} This category investigates the integration of MLLMs with GAI to develop more efficient AD solutions.
\end{enumerate}

\section{LLMs for Autonomous Driving (RQ2)}
Various approaches have been proposed for integrating LLMs with ADS. 
For instance, the authors of \cite{tanahashi2023evaluation} quantitatively evaluated the Spatial-Aware Decision-Making (SADM) and Traffic Rules Satisfaction (TRS) abilities of LLMs. 
Moreover, to implement LLMs within ADS, several strategies have been developed, including methods based on prompt engineering, fine-tuning, RLHF, and GAI.

    \subsection{Prompt Engineering-based Methods}
\begin{figure}[t]
    \centering
    \includegraphics[width = \columnwidth] {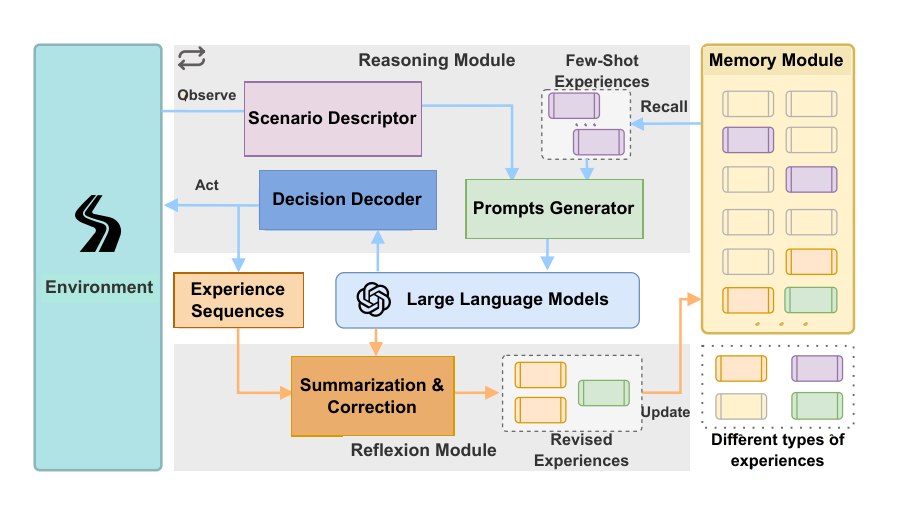}
    \caption{Architecture of DiLu framework \cite{wen2023dilu}.}
    \label{dilu}
\end{figure}

    Prompt engineering uses queries (prompts) to guide the output of LLMs. In this section, we discuss the related work that proposed LLM-based prompt engineering towards AD task provisioning.
Planning and control are critical components that determine the AV's ability to navigate and make decisions in real time. Recent advancements in LLMs and prompt engineering enabled novel approaches to enhance planning and control in ADS. For instance, Zhou \textit{et al.} integrated in \cite{zhou2024context} LLMs with RL to enhance AD agents, making them more efficient. The proposed framework integrated GPT-3.5-turbo with an RL agent based on deep Q-Learning to take driving control actions. The LLM is used as a proxy for reward calculations, i.e., it interprets textual prompts (e.g., task description, objective, and last outcome) to generate reward signals that influence the RL agent’s behavior. Results showed that RL agents guided by LLMs achieved more balanced and human-like behaviors compared to traditional RL agents.
Also, authors of \cite{wen2023dilu} designed the DiLu framework, illustrated in Fig.~\ref{dilu}, that integrated LLM
in ADS to develop four modules, namely an AD simulation environment module, an AD memory module to acquire and save experience from past driving scenarios, a reasoning module that generates reasoning chains and provides AD decisions, and a reflection module to correct the reasoning process for future driving scenarios.  
Authors of \cite{sha2023languagempc} designed ``LanguageMPC'', which is an LLM prompt engineering-based framework, that uses LLMs in reasoning and understanding high-level information with Model Predictive Control (MPC) to execute specific driving actions.

In \cite{azarafza2024hybrid}, the authors proposed a generative driver agent simulation framework to perceive complex traffic scenarios and provide realistic driving maneuvers. Their framework uses LLMs to generate responses based on input prompts, enabling the driver agent to comprehend complex driving scenarios. 
The authors compared the accuracy of LLMs to human-generated ground truth using the CARLA simulator and found that combining detected objects and sensor data in the LLM provides precise brake information. 

\begin{figure*}[t]
    \centering
    \includegraphics[width=0.93\textwidth] {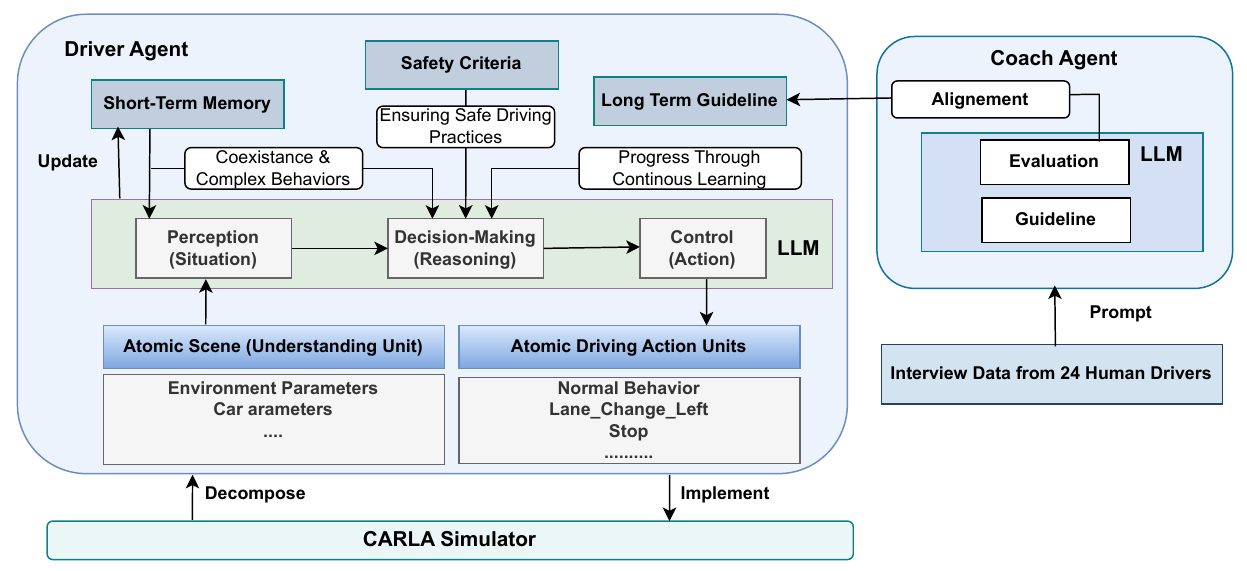}
    \caption{Architecture of the SurrealDriver framework based on \cite{jin2023surrealdriver}.}
    \label{surreal}
\end{figure*}

 The SurrealDriver framework proposed in \cite{jin2023surrealdriver} and illustrated in Fig. \ref{surreal}, is a generative driver agent simulator that uses LLMs. It is composed of three main modules, namely DriverAgent, CoachAgent, and short-term memory, it performs multiple tasks, including perception and control. Within the SurrealDriver framework, the LLM processes the collected environment data parameters to comprehend the driving situation and make decisions guided by predefined requirements and safety rules.  
 The DriverAgent module generates JSON-formatted commands to control the vehicle such as maintaining speed, lane changing, stopping, speeding up, and slowing down.
The CoachAgent module is based on the use of CoT prompts of feedback from 24 drivers to refine the DriverAgent behavior and make it similar to human driving. Finally, the short-term memory stores recent driving behaviors to ensure continuity and consistency in decision-making. Similarly, authors of \cite{miceli2023dialogue} proposed a system that generates self-driving simulation scenarios using natural language dialogue with GPT-4. It allows iterative refinement of scenarios through user-induced language-based instructions and corrections, which are translated by the LLM into executable simulation code.
 
\begin{figure}[t]
    \centering
    \includegraphics[width = 0.5\textwidth] {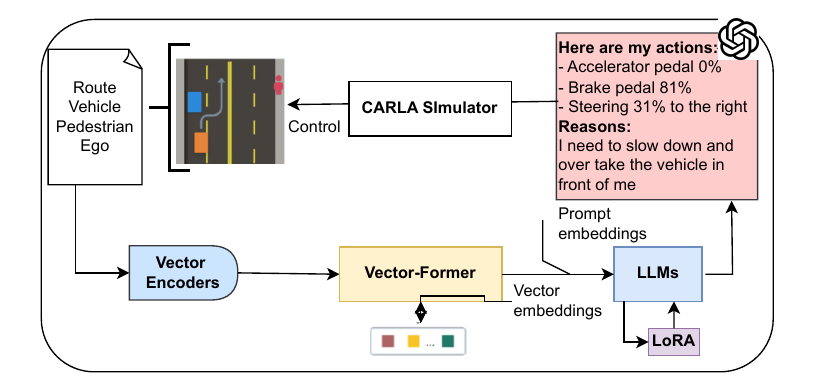}
    \caption{Architecture of LLM-driver framework based on\cite{chen2023driving}.}
    \label{llmdriver}
\end{figure}

LLMs for question-answering focus on designing and optimizing prompts to respond to various questions. 
Accordingly, authors of \cite{chen2023driving} proposed "LLM-driver", shown in Fig. \ref{llmdriver}, a framework that integrates numeric vector modalities into pre-trained LLMs. This method uses object-level 2D scene representations to fuse vector data into a pre-trained LLM with adapters. 
The model can interpret driving situations and generate adequate actions. 
A dataset with 160,000 Q\&A pairs generated by GPT-3.5 from 10,000 driving scenarios was introduced and associated with high-quality control commands, collected from an RL driving agent, to fine-tune the model.
The model’s performance was evaluated in terms of action prediction Mean Absolute Error (MAE), traffic light detection accuracy, and normalized errors of acceleration, brake pressure, and steering.

  \subsection{Fine-Tuning-based Methods}

Wang \textit{et al.} developed in \cite{wang2024drivecot} DriveCoT, an E2E driving dataset using the CARLA simulator. It included complex driving scenarios (e.g., lane-changing and high-speed driving), and incorporated a CoT labeling scheme to provide reasoning processes for driving decisions. 
Also, they designed a DriveCoT-agent model trained on the DriveCoT dataset. The latter showcased strong performances in open-loop and closed-loop evaluations. 

In \cite{peng2024lc}, the authors proposed Lane Change Large Language Model (LC-LLM), an explainable lane change prediction model reconceptualized as a language modeling problem and solved using LLMs. The core of the LC-LLM framework
is a fine-tuned LLM for the lane change and
trajectory prediction tasks for highway AD. Through experiments, it is shown that LC-LLM can accurately predict lane change intentions and trajectories compared to benchmarks based on long Short-Term Memory (LSTM) or Transformers. 


  \subsection{RLHF-based Methods}
  Authors of \cite{yang2024driving} designed driver agents for AD with reasoning and decision-making capabilities that are based on LLM, and aligned with human driving behaviors.
The proposed multi-alignment framework utilizes demonstrations and feedback to align LLM-based driving agents' behaviors with those of humans. The effectiveness of the proposed framework was validated using the CARLA simulator.

In \cite{tian2024enhancing}, a closed-loop framework was proposed to enhance the training and evaluation processes of RL agents for AD, called CRITICAL. It focuses on generating critical driving scenarios to resolve specific learning and performance gaps within the RL agent. The framework included an LLM component to refine and diversify scenario generation based on historical training data and real-world driving knowledge. 
To evaluate the criticality of scenarios, CRITICAL employed surrogate safety measures, e.g., time-to-collision and unified risk index.  


\begin{table*}[t]
    \centering
    \caption{Comparative Study of LLM Approaches for AD}
    \label{compareLLMapproach}
    \begin{tabular}{|p{0.4cm}|p{0.45cm}|p{1.65cm}|p{4cm}|p{4.5cm}|p{4.5cm}|}
        \hline
        \textbf{Year}&\textbf{Ref.} & \textbf{Model} & \textbf{Used Datasets and Tools} & \textbf{Trainable Modules} & \textbf{AD Services} \\
        \hline
                 {\quad 2023} &
         \cite{tanahashi2023evaluation}   & GPT-3.5-turbo-1106 & HighwayEnv & RL agent integrating LLM  & Human-like driving behavior; Dynamic adaptation to driving conditions; Customized driving styles.
         \\
        \hline 
         \quad 2023 & \cite{sha2023languagempc}  & ChatGPT-3.5 & IdSim & MPC; RL agent & Single-vehicle decision-making; Multi-vehicle coordination; Driving behavior modulation  \\
            \hline

        \quad 2023 & {\cite{jin2023surrealdriver}} & GPT-3; GPT-4  & Driving Behavior Data, simulation data generated with CARLA; CARLA simulator, NLP library & DriverAgent (Perception, control, and decision-making); CoachAgent (Driving scenario generation and evaluation) & ADS simulation and testing; Safety assurance; Real-time monitoring
             \\
         \hline 
        \quad 2023 & {\cite{chen2023driving} }  
               & GPT-3.5 & 160k QA driving pairs dataset and Control Commands Dataset (15 virtual environments); 2D driving simulator & PPO-based RL agent; LoRA modules & Perception, action prediction, and driving Q\&A
            \\  \hline

        \quad 2024 & {\cite{zhou2024context}}   & GPT-3; GPT-4; BERT; T5  & CitySim dataset; Highway-env,  graph representation and box plots & Few-shot learning model & Closed-loop driving task. 
            \\
            \hline

        \quad 2024 & {\cite{azarafza2024hybrid}}& GPT-4 & CARLA simulator and YOLOv8 object detector & Common sense knowledge; Common sense reasoning & Object detection and localization in in diverse weather conditions.
             \\
            \hline
            
         \quad 2024 & {\cite{wang2024drivecot}}  &  Unspecified  & Private dataset generated with CARLA; CARLA simulator  & Unspecified  & Interpretability of E2E ADS: Perception, planning, prediction, and reasoning
        \\
            \hline
            
    \quad 2024 & {\cite{peng2024lc} }  &  LLaMA-7B; LLaMA-13B; LLaMA-70B & highD dataset; LoRA, & Supervised fine-tuning &  Lane change intention prediction; Explainable predictions \\
            \hline

     \quad 2024 & {\cite{yang2024driving} } & GPT-4  & Private dataset from human drivers; CARLA simulator, NLP library, RL library &  Language understanding; Behavioral cloning; RL agent; Decision-making & Behavioral alignment; Human-in-the-loop system
        \\
            \hline
            
    \quad 2024 & {\cite{tian2024enhancing}} & Mistral-7B-Instruct  & HighD Dataset; HighwayEnv Simulation, and LongChain & Scenario generation; PPO-based RL agent; closed-loop feedback & AD scenario generation; Dynamic scenario modification; Safety measures  analysis
             \\
            \hline
      \quad 2024 & {\cite{zhao2024drivedreamer}} & GPT-3.5 & Trajectory-to-HDMap Dataset and Multi-View Video Dataset; Python libraries & Function Library; Embedding HDMaps and 3D Boxes; UniMVM module &  Perception; Video generation of driving scenarios; AD scenario simulation 
             \\\hline
        
        \end{tabular}
      \end{table*}

\subsection{LLM and GAI-based Method}
\begin{figure*}[t]
    \centering
    \includegraphics[width = 0.9\textwidth] {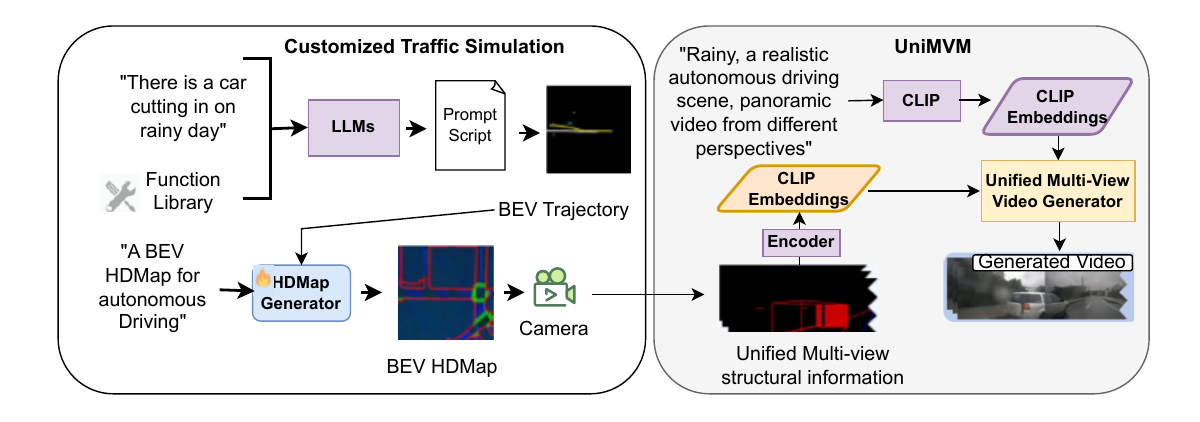}
    \caption{Architecture of DriverDream-2 framework  \cite{zhao2024drivedreamer}.}
    \label{dream2}
\end{figure*}

The DriveDreamer-2 framework proposed in \cite{zhao2024drivedreamer}, which is an extension of DriveDreamer  \cite{wang2023drivedreamer}, was designed to generate user-customized synthetic and realistic multi-view driving videos used for training, testing and validation of ADS efficiency. Specifically, it generates 
driving videos by combining structured conditions (e.g., HD-Maps and 3D boxes) with image features. The system uses encoders to embed HD-Maps, 3D boxes, and image frames into latent space
features, which are then processed to produce the final videos. Then, using the Unified Multi-View Model (UniMVM), the spatial and temporal coherence of generated videos is enhanced. The architecture of DriverDream-2 is presented in Fig.~\ref{dream2}


In Table \ref{compareLLMapproach}, we present a comparative study of the discussed works above.

\section{VLMs for Autonomous Driving (RQ3)}

\subsection{Prompt Engineering-based Method}

\begin{figure} [t]
    \centering
    \includegraphics[width = 0.48\textwidth] {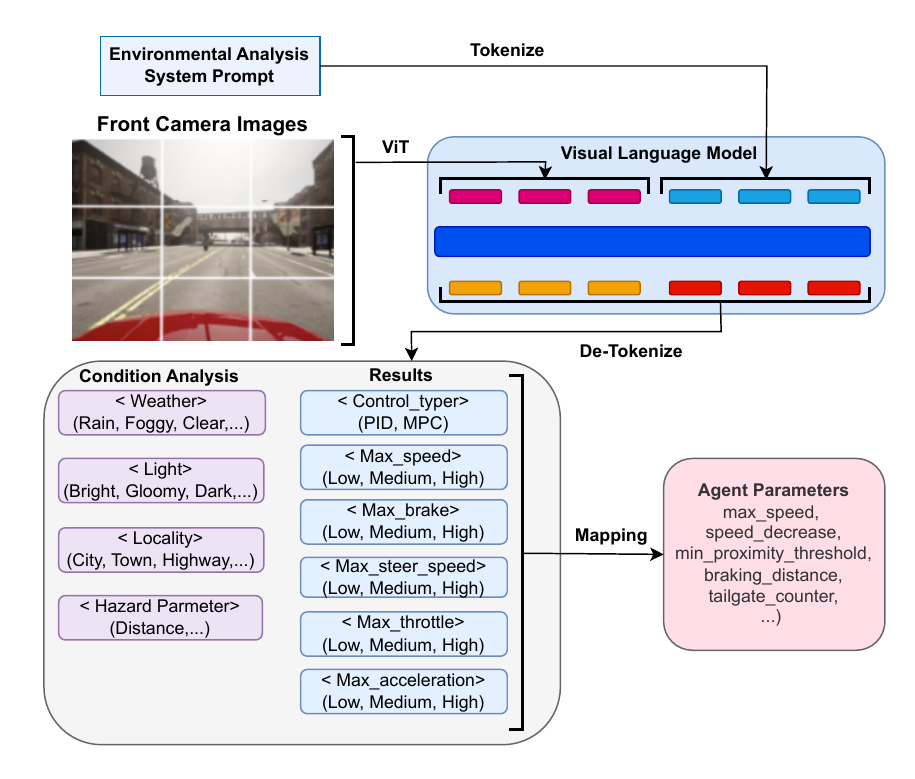}
    \caption{Architecture of Co-driver framework  based on \cite{guo2024co}}
    \label{codriver}
\end{figure}

In \cite{guo2024co}, Guo \textit{et al.} proposed the Co-driver framework, based on ViT, illustrating the potential of prompt engineering in planning and control, and trajectory prediction tasks. Co-driver employs prompt engineering to understand visual inputs and generate driving instructions, which are then fed to a Deep RL (DRL) agent for driving actions. 
Also, it predicts trajectories through the integration of map contexts and past vehicle positions using scene encoding and path classification. 
The Co-driver architecture is presented in Fig.~\ref{codriver}.
 \subsection{Fine-Tuning-based Methods} 
 Perception in ADS is crucial for interpreting and understanding the vehicle's surroundings, thus accurately detecting and recognizing objects, pedestrians, and road conditions. This yields informed decisions in real-time by the ADS, thus improving the overall safety and performance. 

\begin{figure*}[t]
    \centering
    \includegraphics[width = 0.85\textwidth] {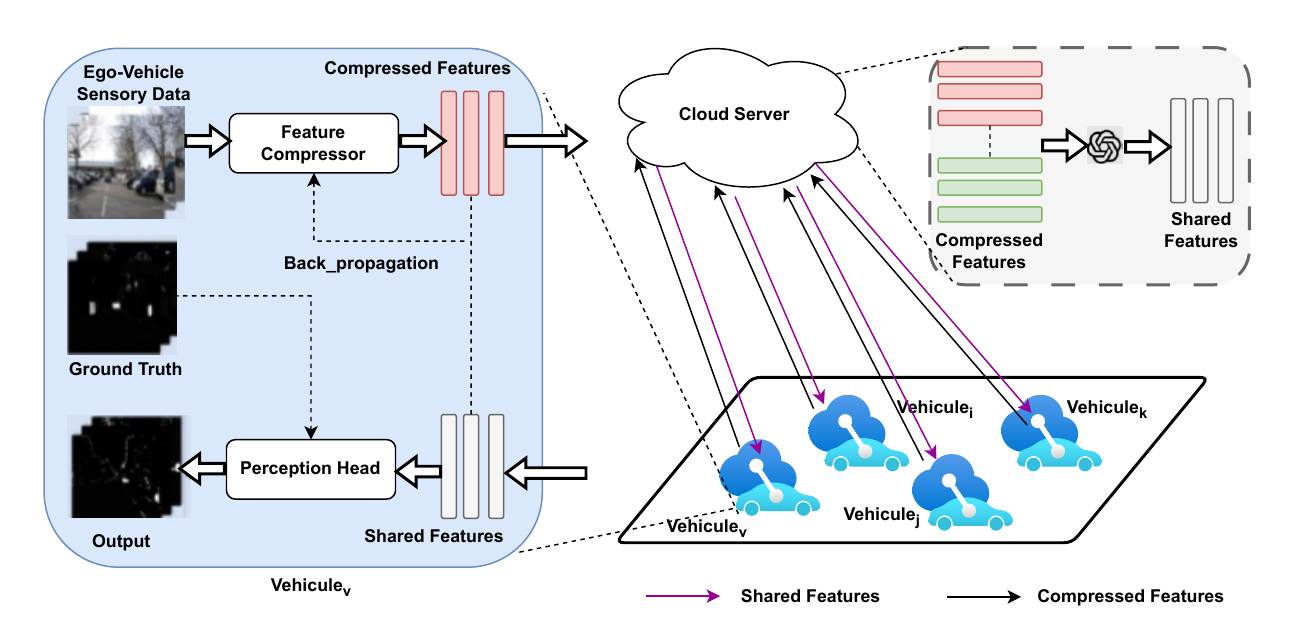}
    \caption{Architecture of pFedLVM framework based on \cite{kou2024pfedlvm}.}
    \label{pfelvm}
\end{figure*}

 In \cite{tian2024drivevlm}, Tian \textit{et al.} proposed DriveVLM and 
 DriveVLM-Dual frameworks. DriveVLM uses a ViT encoder to process images and extract visual features. Then, a text tokenizer and an LLM, supported by CoT, process textual data and generate scene descriptions and driving decisions. Finally, CoT decisions are converted into vehicle control commands for execution. 
 DriveVLM focused on the scene understanding and planning tasks, while DriveVLM-Dual extended the work to integrate VLMs with conventional AD methods, thus strengthening the spatial understanding of the driving environment and speeding real-time inference.
Moreover, the corner Case Object Detection and Analysis for Large Models (CODA-LM) framework was proposed in \cite{li2024automated} to evaluate VLMs' performances in complex driving scenarios. 
It provides a detailed analysis of the VLMs' strengths and weaknesses when handling corner cases, thus providing insights into the areas that need further investigation to enable AD.
Finally, Kou \textit{et al.} proposed in \cite{kou2024pfedlvm} the Personalized Federated Learning Large Vision Model (PFedLVM) framework aiming to improve perception by leveraging Federated Learning (FL) and personalization. Specifically, LVM is deployed only within the aggregation server, the latent feature-based FL is used to exchange compressed feature maps, while personalized learning ensures that each AV model learns from the others but preserves its unique characteristics. Its architecture is presented in Fig. \ref{pfelvm}.

\begin{figure}[t]
    \centering
    \includegraphics[trim={1cm 1cm 0.5cm 0cm},clip,width = \columnwidth] {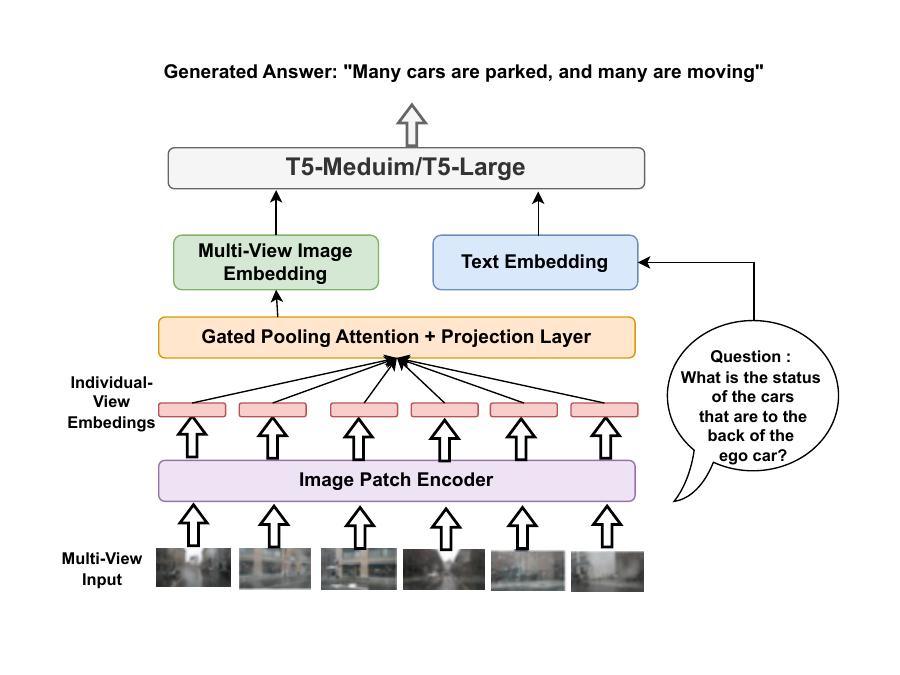}
    \caption{Architecture of EM-VLM4AD framework \cite{gopalkrishnan2024multi}.}
    \label{emvlm}
\end{figure}

Authors of \cite{gopalkrishnan2024multi} proposed ``Efficient, Lightweight Multi-Frame Vision Language Model for Visual Question Answering in Autonomous Driving'', a.k.a., EM-VLM4AD, a VLM model focusing on performing Q\&A tasks for AD.  
To produce interpretable responses, EM-VLM4AD integrates multiple camera views into a unified visual representation and combines it with text embeddings generated with ViT. Specifically, it uses a fine-tuned T5-Medium model and an 8-bit quantized T5-Large model fine-tuned via LoRA to align with the concatenated multi-view image and text embeddings. The EM-VLM4AD architecture is shown in  Fig.~\ref{emvlm}.





In Table \ref{comparevLMapproach}, we present a comparative study of the
discussed works in Section VIII.
 
\begin{table*}
    \centering
    \caption{Comparative Study of VLM Approaches for AD}
    \label{comparevLMapproach}
    \begin{tabular}
    {|p{0.4cm}|p{0.45cm}|p{1.65cm}|p{4cm}|p{4.5cm}|p{4.5cm}|}
        \hline
        \textbf{Year}&\textbf{Ref.} & \textbf{Model} & \textbf{Used Datasets and Tools} & \textbf{Trainable Modules} & \textbf{AD Services} \\
        \hline
        \quad 2024 & \cite{guo2024co} & Qwen-VL-9.6B 
        &  Customized dataset with image sets and corresponding prompts; CARLA simulator and ROS2 & Qwen-VL fine-tuning using Quantized Low-Rank Adaptation (QLoRA) & Adjustable driving behaviors; Trajectory and lane prediction; Planning and control
        \\
        \hline
        \quad 2024 & {\cite{tian2024drivevlm}} & Qwen-VL &  SUP-AD dataset and NuScenes &	ViT encoder; LLM	& Scene understanding; Motion planning; Decision-making; Trajectory planning
        \\
        \hline
        
        \quad 2024 & {\cite{li2024automated}} & Flamingo & Ego4D, Waymo Open Dataset, NuScenes, and BDD100K;  Cityscapes, OpenCV, PyTorch, COCO API, and Detectron2  &  ViT encoder; CLIP & Object detection; Scene understanding; Path planning; Obstacle avoidance.
        \\
        \hline
        \quad 2024 & {\cite{kou2024pfedlvm}} & ViT & Cityscapes and CamVid & Feature extraction; Compression; Backpropagation  & Object detection; Semantic segmentation; Vehicle-specific behavior modeling
        \\
        \hline
        
        \quad 2024 &  {\cite{gopalkrishnan2024multi}} & T5-Medium; T5-Large; ViT & DriveLM dataset; LoRA,  &  Attention and Projection; 
        LLM fine-tuning &  Question-answering
          \\
         \hline    
    \end{tabular}
       \end{table*}
\section{MLLMs for Autonomous Driving (RQ4)}

\subsection{Prompt Engineering-based Methods}

\begin{figure*}[t]
    \centering
    \includegraphics[width =  0.95\textwidth] {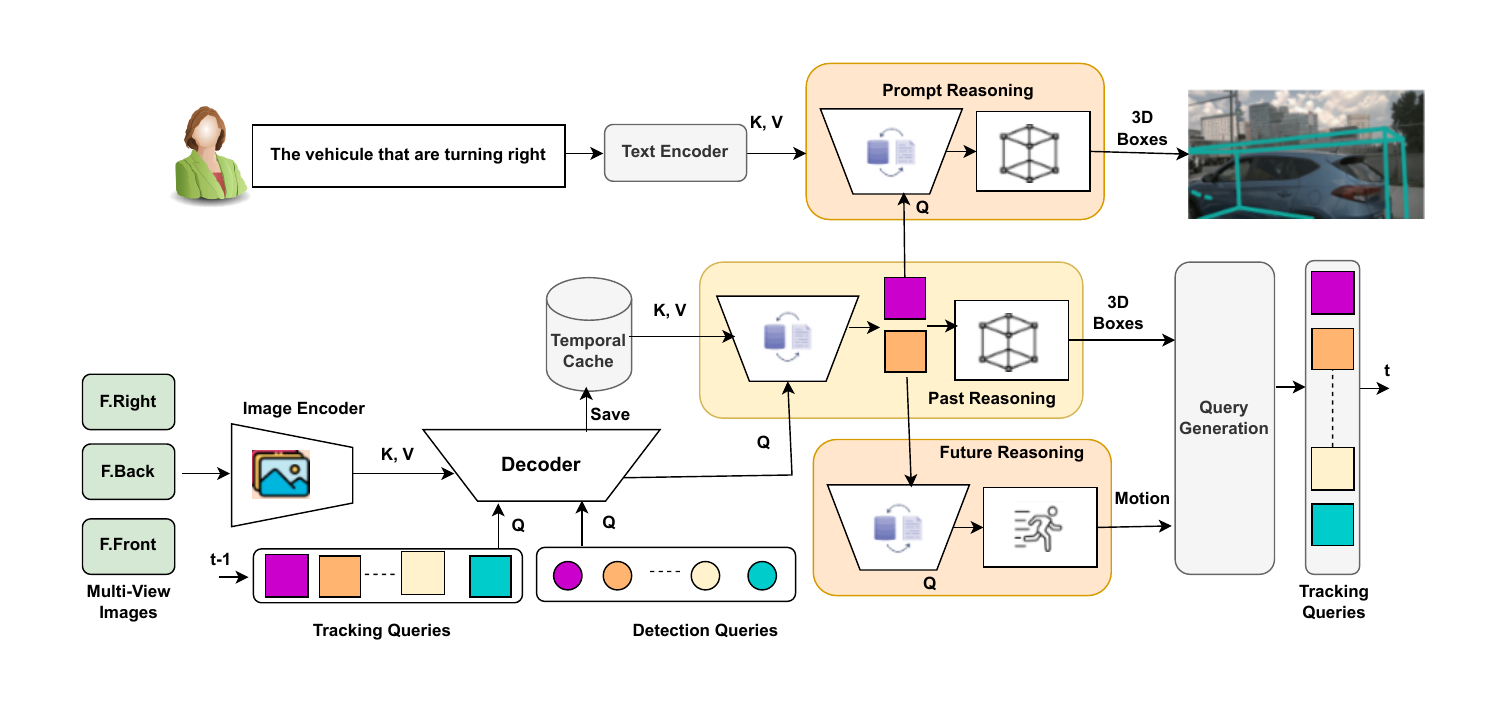}
    \caption{Architecture of PromptTrack framework \cite{wu2023language}.}
    \label{promttrack}
\end{figure*}

In \cite{wu2023language}, Wu \textit{et al.} proposed a new large-scale language prompt dataset for driving scenes, called ``NuPrompt'', specializing in 3D objects  and is built on NuScenes dataset \cite{caesar2020NuScenes}
for multi-view 3D object detection.        
Also, they introduced an approach to 3D object detection and tracking by integrating cross-modal features within prompt reasoning, called ``PromptTrack''. This approach outperformed traditional object detection methods, due to its novel fusion of multi-modal inputs.
The architecture of PromptTrack is illustrated in Fig.~\ref{promttrack}, where the Transformer decoder processes each frame's visual attributes and inquiries to generate decoded questions. Also, the past reasoning module improves and refines tracking based on historical queries, whereas the future reasoning module facilitates cross-frame query propagation. Finally, the prompt reasoning branch predicts prompt-related tracks. 
Authors of \cite{ding2023hilm} proposed ``High-Resolution Understanding in MLLMs for Autonomous Driving'', a.k.a., HiLM-D, as an efficient technique to incorporate a high-resolution information into MLLMs for risk object localization and intention prediction. Through experiments, the authors proved the superiority of HiLM-D in localizing obstacles and predicting vehicles' intentions compared to two benchmarks, namely eP-ALM \cite{shukor2023ep} and Video-LLaMA \cite{zhang2023video}.
Finally, RAG-driver is a recent framework
based on prompt engineering that generates driving explanations where the MLLM is responding to queries about driving scenarios \cite{yuan2024rag}.   
It leverages a retrieval engine that looks for similar driving experiences to the one under analysis, then the MLLM processes both the current query and the in-context learning samples that have been retrieved in its memory to provide action explanation, justification, and next control signal prediction. 

\subsection{Fine-Tuning-based Methods}
\begin{figure*}[t]
    \centering
    \includegraphics [width =  \textwidth] {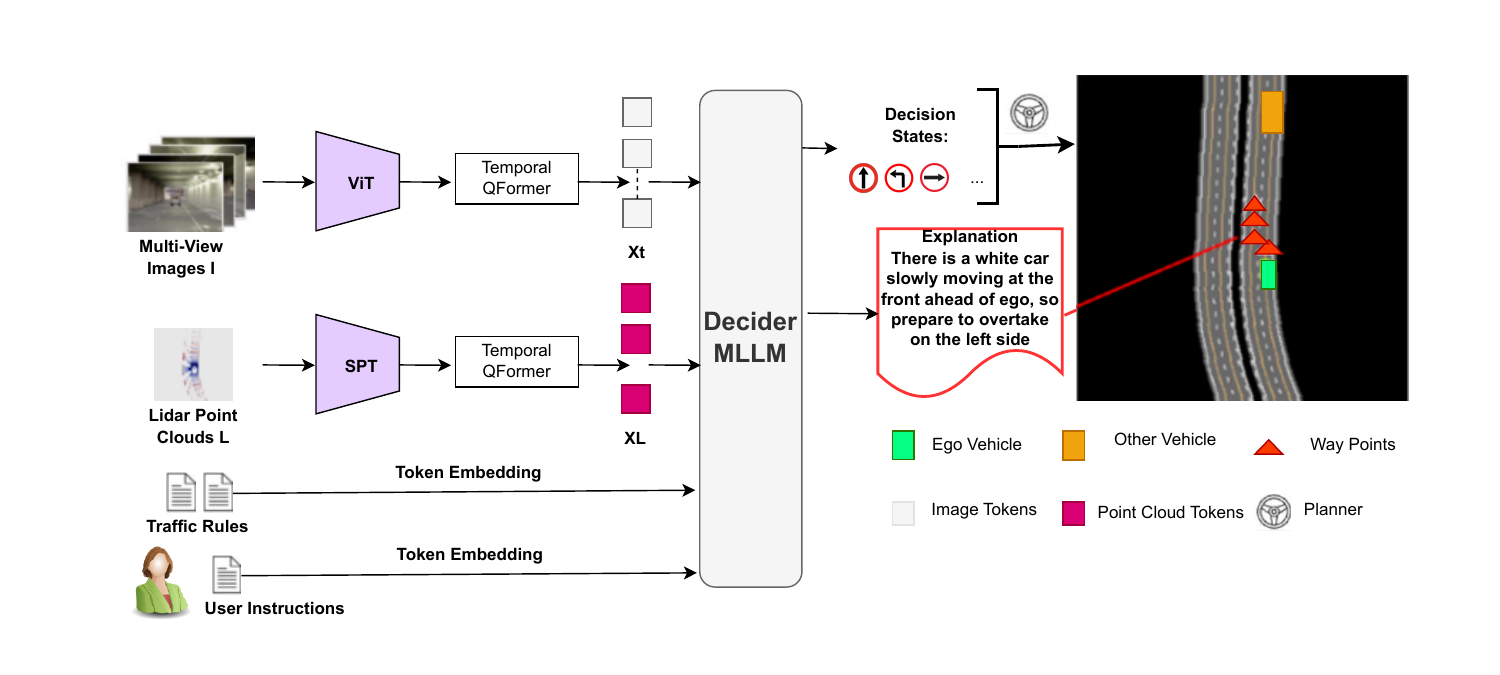}
    \caption{Architecture of DriveMLM framework \cite{wang2023drivemlm}.}
    \label{DriveMLM}
\end{figure*}

In \cite{wang2023drivemlm}, the authors proposed DriveMLM, an LLM-based framework performing close-loop AD in realistic simulators. Indeed, it integrates three modules: (1) the ``Behavioral Planning'' module that utilizes MLLM to incorporate user commands, driving rules, and sensor inputs for driving decisions and explanations, (2) the ``Behavioral Planning States Alignment'', which aligns LLM’s linguistic decision outputs with the behavioral planning module, and (3) the ``Data Engine'' which collects data with decision states and explanations for model training and evaluation. According to the authors, DriveMLM improves driving scores compared to Apollo baseline, on the ``CARLA Town05 Long'' benchmark. The DriveMLM framework is illustrated in Fig.~\ref{DriveMLM}.
 Moreover, Han \textit{et al.} designed in \cite{han2024dme} an ADS named  Decision-Making and Execution-driver (DME-Driver) that generates NLP-based driving decisions based on correlation vehicle status and visual inputs, which are then converted into control commands. It is based on a pre-trained CLIP visual encoder to convert visual information into feature tokens, and a text tokenizer to encode prompt inputs and current status information. The latter are processed using LLaMA 2 model. It has been shown that DME-Driver outperforms GPT-4V in terms of accuracy across several tasks, including gaze (85.2\% vs. 75.3\% for GPT-4V), scene understanding (86.5\% vs. 79.2\% for GPT-4V), and logic (80.3\% vs. 65.4\% for GPT-4V).
DriVLMe, presented in \cite{huangdrivlme}, is a fine-tuning MLLM for trajectory planning. 
It is a video-language-model-based AD agent facilitating communication between humans and AVs that perceive the environment and navigate. Specifically, it learns from embodied experiences
in a simulated environment and social experiences from real human dialogue to find the shortest paths from the agent’s current location to the destination specified by the MLLM.
In addition, Liao \textit{et al.} proposed in \cite{liao2024vlm2scene} VLM2Scene, a fine-tuning method for AD perception. 
It shifts from traditional point-level contrastive learning to region-level learning to address the inherent sparsity and noise in LiDAR point clouds. To do so, it leverages region masks derived from the Segment Anything Model (SAM). VLM2Scene fine-tunes the learning process to enhance the model's perception accuracy and robustness by introducing a semantic-filtered region-learning and a region-semantic assignment strategy. Similarly, \cite{liang2024aide} designs the Automatic Data Engine (AIDE) that automates data labeling, model training, and driving scenario generation for model evaluation, aiming to improve AV perception models continuously.  
From the multi-tasking perspective, Ding \textit{et al.} presented in \cite{ding2024holistic} their BEV-InMLLM framework, for which they introduced NuInstruct, a novel dataset including 91,000 multi-view video-QA pairs across 17 subtasks. One of the main contributions of this work consists of injecting Bird's-Eye-View (BEV) data into MLLM to fine-tune the models. The proposed BEV-InMLLM approach uses a specialized multi-view Q-Former to handle multi-view video inputs and capture temporal appearance information across different views, thus accurately localizing objects and estimating distances. Furthermore, the framework OmniDrive features a novel 3D vision-language model that utilizes sparse queries to extract and compress visual representations into 3D before inputting them into a language model and a comprehensive Visual Question Answering (VQA) module that includes traffic regulation, scene description, 3D grounding, decision-making, planning, and counterfactual reasoning \cite{wang2024omnidrive}. Based on the experiments, OmniDrive demonstrated high reasoning and planning capabilities in complex 3D scenes.
Finally, DriveGPT4 proposed fine-tuned MLLMs for question-answering in ADS \cite{xu2023drivegpt4}. DriveGPT4 is an interpretable E2E ADS that processes both multi-frame video inputs
and textual queries to explain taken AD actions to the human user. 

\subsection{RLHF-based Methods}

\begin{figure*}[t]
    \centering
    \includegraphics[width =0.9 \textwidth] {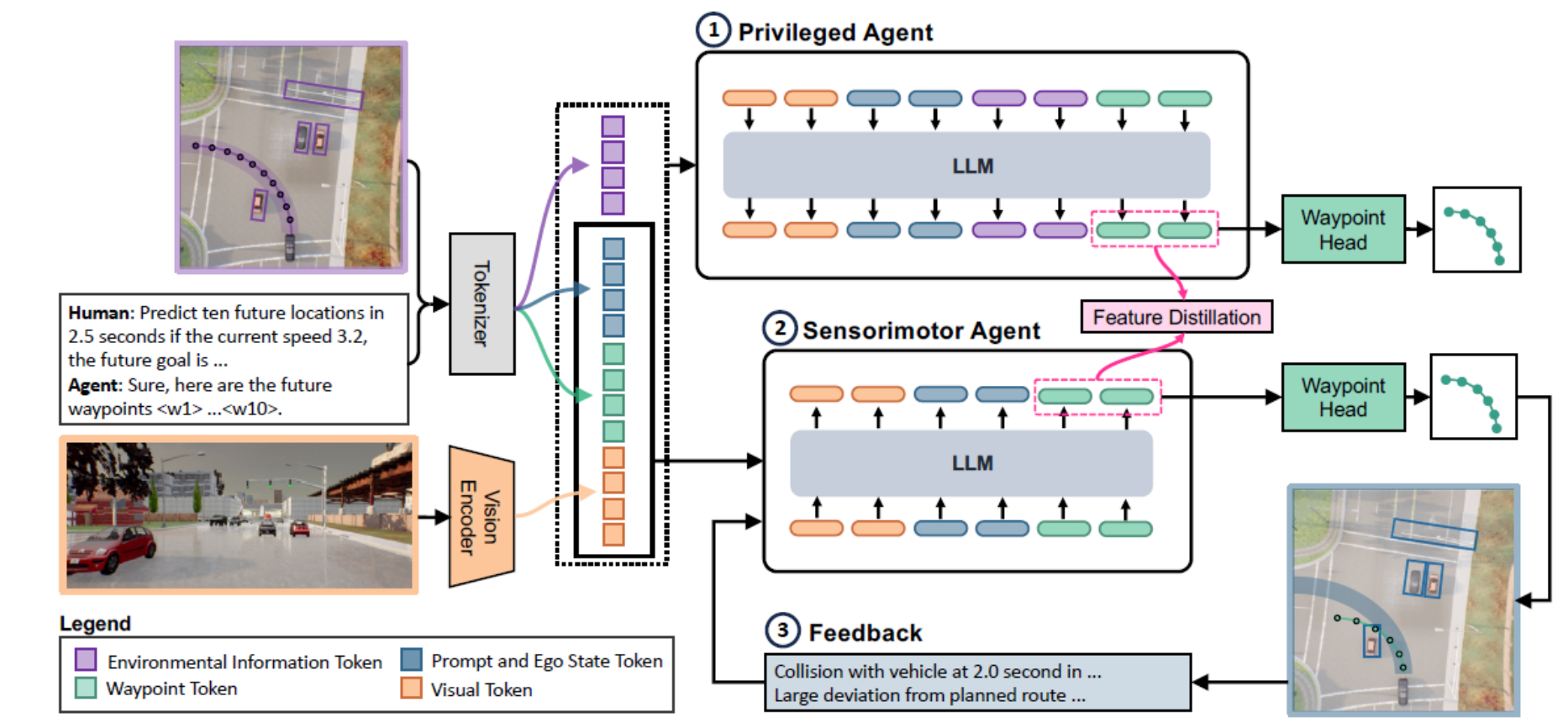}
    \caption{Architecture of the Feedback-Guided-AD \cite{rayfeedback}.}
    \label{fedarchi}
\end{figure*}

Mao \textit{et al.} presented in \cite{mao2023language} Agent-driver, an MLLM-based intelligent agent for AD that integrates: (1) a versatile tool library for dynamic perception and prediction, (2) a cognitive memory for human knowledge, and (3) a reasoning engine that emulates human decision-making using CoT technique. 
Through experimentation, Agent-driver is proven to outperform state-of-the-art AD methods in terms of interpretability and few-shot learning. Also, authors in \cite{rayfeedback} proposed a feedback-guided E2E sensorimotor driving agent with MLLM to provide a language interface for user control and refinement. The objective is to train the sensorimotor agent to map front camera images and AV's state information encoded as language tokens and predict a set of future waypoints. The training phase involves two stages: 1) The privileged agent takes ground truth environmental information and provides rich supervision for training the sensorimotor agent, and 2) the sensorimotor agent is fine-tuned with prompt-based feedback to enable efficient failure reasoning. Simulation results demonstrated the efficiency of this method.
Further architectural details of this framework are illustrated in Fig. \ref{fedarchi}.

\subsection{MLLM and GAI-based Methods}

Authors of \cite{jia2023adriver} introduced the notion of the interleaved vision-action pair, which unifies the format of visual features and control signals, leading to the general world model, called ADriver-I. 
The latter can predict the control signal of the current frame based on the vision-action pairs as inputs. Then, it predicts and generates future frames based on the 
generated control signals with the vision-action pairs. 


Finally, Table \ref{compareMLLMapproach} presents a comparative study of the discussed works in Section IX.

\begin{table*}[ht!]
    \centering
    \caption{Comparative Study of MLLM Approaches for AD}
    \label{compareMLLMapproach}
    \begin{tabular}{|p{0.4cm}|p{0.45cm}|p{1.7cm}|p{3.95cm}|p{4.5cm}|p{4.5cm}|}
        \hline
      \centering  \textbf{Year}&\textbf{Ref.} & \textbf{Model} & \textbf{Used Datasets and Tools} & \textbf{Trainable Modules} & \textbf{AD Services} \\
        \hline
     \centering {\quad 2023} & \centering {\cite{xu2023drivegpt4}}   &   LlaMA-2; LlaVA; Valley  & CC3M and WebVid-2M datasets & Video tokenizer (based on Valley); CLIP encoder; Mix-fine-tune module; Text tokenizer/de-tokenizer & Question-answering task; Control    
      \\
        \hline

    \centering \quad 2023 & {\cite{ding2023hilm}} & BLiP-2; Q-Former; MiniGPT-4 & DRAMA dataset; Pytorch libraries & Visual encoder; Query detection; Incorporation module ST-adapter & Risky object detection 
      \\
        \hline
    \centering \quad 2023 & {\cite{wang2023drivemlm}}   & LlaMA-7B; Fine-tuned GD-MAE on ONCE & CARLA-generated dataset; CARLA simulator  &  Visual encoder: ViT-g/14 from EVA-CLIP; Q-Former; Token embedding  & Perception; Reasoning; Question-answering; Planning
      \\
        \hline
      \centering  \quad 2023 & {\cite{mao2023language}} & Llama-2-7B; GPT-3.5-turbo-1106; GPT-3.5-turbo-0613 & NuScenes dataset & Detection; Prediction; Occupancy; Mapping & Task planning; Motion planning; Reasoning; Driving actions\\
         \hline
        
 \centering  \quad 2023 & {\cite{jia2023adriver}}    & Diffusion Models;  Vicuna-7B-1.5; LlaVA-7B-1.5  & NuScenes and private datasets   & ViT; CLIP-ViT-Large; Multi-layer perceptrons  & Future scene prediction; Action prediction      \\
        \hline      

 \centering \quad 2024 & {\cite{yuan2024rag}} & Small MLLM-7B & BDD-X and Spoken-SAX datasets  & Clip4clip; VIT-B/32 & Q\&A for planning; Perception Control   \\
 \hline

\centering \quad 2024 & {\cite{han2024dme}} & LlaMA-2; CLIP  & HBD dataset & Perception modules (TrackFormer, MapFormer, MotionFormer, and OccFormer); Planning module & Decision-making; Scene understanding; Motion prediction; Vehicle control 
        \\
        \hline

\centering \quad 2024 & { \cite{huangdrivlme}} &  Vicuna-7B v1.1 
& For closed-loop: CARLA simulator;  For open-loop: Situated dialogue navigation and BDD-X datasets  & Video tokenizer; Text tokenizer; LLM backbone; CLIP encoder; Route planning  & Dialogue tasks; Route planning \\
        \hline

\centering \quad 2024 & {\cite{liao2024vlm2scene}}& CLIP; BLIP-2; SAM &  NuScenes, KITTI, and Waymo datasets  & 3D network (E3D); Region semantic concordance regularization; Region caption prompts & Perception; Contextual awareness\\
        \hline
\centering \quad 2024 & {\cite{liang2024aide}} & Dense captioning models; OWL-ViT; CLIP; Otter; ChatGPT	& LVIS dataset; OWL-v2 & Pseudo-labeling models; Fine-tuning models	& Perception; Object detection
         \\
        \hline

\centering \quad 2024 & {\cite{ding2024holistic}}& BLIP-2; MiniGPT4; Video Llama; BEV Extractor  & NuInstruct dataset & Q-Former; Injection module; & Perception; Risk assessment; Prediction; Planning.\\
        \hline

\quad 2024 & \centering {\cite{wang2024omnidrive}} & BLIP-2; LlaVa 1.5; GPT-4V; Lora & NuScenes dataset  & Depth-first algorithm; Q-Former 3D MLLM & VQA generation; Perception-action alignment; Decision-making; Planning \\
 \hline
    
 \centering    \quad 2024 & \centering {\cite{rayfeedback} }  & LLaVA-7B; LlaMA  & NuScenes dataset; CARLA simulator     & CLIP; Vit; Token prediction; Vision encoder; Language encoder; Waypoint prediction & Future location prediction
      \\
        \hline

 \centering  \quad 2024 & \centering {\cite{wei2024editable}}   & 7 LLM agents, each for a specific task  &  Waymo dataset &  McNeRF and McLight for background and foreground rendering  & Editing 3D driving scenes
    \\
    \hline
    \end{tabular}
      \end{table*}

\section{Datasets \& Simulators for ADS (RQ5)}
\subsection{Datasets}
The deployed datasets play a pivotal role in developing large-scale models for ADS. Indeed, ADS relies heavily on extensive datasets with various driving conditions, including changing lighting, weather patterns, and road environments. Used datasets should be meticulously annotated to replicate the real-world environment for training and validation purposes. 

Recently, several datasets have been designed for AD, such as NuInstruct \cite{ding2024holistic}, DRAMA \cite{malla2023drama}, and NuScenes-MQA \cite{inoue2024NuScenes}. Indeed,  NuInstruct is a language-driven dataset with 91,000 multi-view video Q\&A pairs corresponding to 17 driving subtasks \cite{ding2024holistic}. It was created by automating instruction-response pairings through a logical sequence similar to the human driving process, i.e., it starts by detecting nearby objects (Perception), forecasts their behaviors (Prediction), assesses their potential danger (Risk prediction), and plans a safe route with reasoning (Planning). To ensure high-quality data samples, each instruction-response pair is validated by humans or GPT-4. 
Also, the ``Driving Risk Assessment Mechanism with A captioning module'', a.k.a., DRAMA, includes 17,785 interactive driving scenarios collected in Tokyo, Japan \cite{malla2023drama}. Each scenario is 2 seconds long, yielding 91 hours of annotated video footage. Video samples have been collected using the SEKONIX SF332X-10X and GoPRO Hero 7 cameras in a real-world urban environment and specific to braking responses to external events. These samples are synchronized with controller area network signals and IMU information. DRAMA is relevant to training XLMs on risk assessment, object detection, and decision-making in dynamic driving environments. 
Finally, NuScenes-MQA is based on NuScenes, which provides full-sentence responses and facilitates the simultaneous evaluation of a model’s capabilities in sentence generation and visual question-answering \cite{inoue2024NuScenes}. To create the Q\&A dataset, the authors employed rich annotations as ground truth from NuScenes. The dataset includes 1,459,933 annotations covering various aspects: (1) Specific object presence, (2) objects, and (3) relative location to the vehicle. Table \ref{tab:datasets} compares the datasets from the task, information, and size perspectives. 
\begin{table*}[t]
\centering
\caption{Comparison of ADS datasets}
\label{tab:datasets}
\begin{tabular}{|c| c c c c| c c c c c c |c|}
\hline
 \textbf{Dataset} & \multicolumn{4}{c}{\textbf{Supporeted Tasks}} & \multicolumn{6}{c}{ \textbf{Included information}} & \textbf{Size} \\
 \hline
 & \textbf{Percep.} & \textbf{Predict.} & \textbf{Risk assess.} & \textbf{Plann.} & \textbf{Multi-view} & \textbf{Tempo.} & \textbf{Multi-obj.}  & \textbf{Dist.} & \textbf{Location} & \textbf{Road} &  \\
\hline
NuInstruct &  $\checkmark$ & $\checkmark$ & $\checkmark$ & $\checkmark$ & $\checkmark$ & $\checkmark$ & $\checkmark$ & $\checkmark$ & $\checkmark$ & $\checkmark$ & 91k    \\
\hline
DRAMA &  $\checkmark$ & $\checkmark$ & $\checkmark$ & $\times$ & $\checkmark$ & $\times$ & $\checkmark$ & $\times$ & $\checkmark$ & $\times$ & 100k    \\
\hline
NuScenes-MQA &  $\checkmark$ & $\checkmark$ & $\checkmark$ & $\checkmark$ & $\checkmark$ & $\checkmark$ & $\checkmark$ & $\checkmark$ & $\checkmark$ & $\checkmark$ & 1.4M    \\
\hline
\end{tabular}
\end{table*}

\subsection{Simulators}
\begin{figure}[t]
    \centering
    \includegraphics[width = 0.5\textwidth]{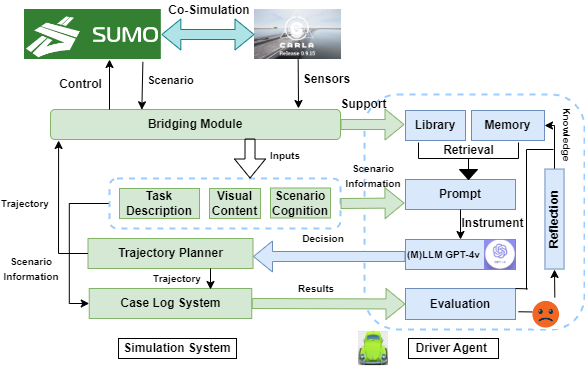}
    \caption{Architecture of LimSim+ \cite{fu2024limsim++}.}
    \label{limsim}
\end{figure}

The complexity and criticality of AD necessitate rigorous evaluation and benchmarking. We review here relevant tools and platforms created for this purpose. First, LimSim++ has been recently proposed as an advanced closed-loop simulation platform for AD \cite{fu2024limsim++}.   
It includes detailed simulations of traffic flow, traffic control, road infrastructure, and environmental conditions, thus enabling robust evaluation of AD performances. Users can use LimSim++ in different ways: 1) Prompt engineering, e.g., a user can create appropriate scenario descriptions and prompt cues for custom scenarios to facilitate the use of MLLMs for vehicle control, 2) model evaluation, e.g., MLLM for AD performance evaluation, and 3) ADS framework improvement, through modification of the ADS sub-modules in closed-loop mode. LimSim++ is composed of three main modules as follows: (1) an information integration module that feeds in scenarios provided by Simulation of Urban Mobility(SUMO) and visual contents from the CARLA simulator, (2) an MLLM prompt engine to understand scenarios and tasks, and (3) a continuous learning module, which allows the driver agent make behavioral decisions. The architecture of LimSim++ is presented in Fig. \ref{limsim}. Moreover, ChatSim is an AD scene simulation that has drawn attention due to its significant potential to produce accurate data for driving scenarios \cite{wei2024editable}. It allows the generation and editing of realistic and customized 3D driving scenes using collaborative LLM. 


\section{Open issues and Future Directions (RQ6)}
Despite the great progress and advancements in XLMs for ADS, their deployment in real-world systems is still facing critical challenges.

\subsection{New Datasets for XLM-assisted ADS}
The XLMs that will be used to enable ADS must synthesize and interpret inputs from multiple modalities, including 3D point clouds, panoramic images, and HD map annotations. Current datasets are limited in scale, quality, and diversity to achieve precise ADS functionalities. 
Most multimodal LLMs, such as GPT-4V, have been pre-trained on open-source datasets that include driving and traffic scenes.
However, the vision-language datasets derived from NuScenes, for instance, do not provide a sufficiently robust benchmark for visual-language understanding in AD contexts. Hence, there is an urgent need to create extensive, diverse, and scalable datasets covering any traffic or driving situation, particularly critical and rare events. In addition, such new datasets should be accompanied by high-quality annotations, e.g., object labels and semantic information, that help in understanding complex scenes and increase the training and evaluation precision. 

\subsection{Mitigating XLM Hallucination}
The hallucination of XLMs refers to the phenomenon where the outputs (e.g., generated text response) are inconsistent with the corresponding visual content, which might be critical in the AD context. Recently, new approaches have been proposed to address hallucination from data, model, training, and inference \cite{bai2024hallucination}. Nevertheless, more in-depth work is needed to mitigate the hallucination effect. Improvement directions would encompass diversifying the training datasets and enhancing their quality, developing systems that explicitly enforce consistency between modalities during training and inference, proposing novel XLM models that natively reduce the hallucination events, e.g., using RLHF, and finally designing simulators and/or testbeds for hallucination assessment.   
\subsection{Enabling XLMs on Resource-limited Hardware}
Given the complexity of XLM methods, it is difficult to deploy them on capacity-limited hardware. 
To bypass this issue, several strategies can be developed. For instance, FL techniques can be leveraged to train models across distributed data sources, e.g., AVs, without requiring centralized data, processing, and storage. Until today, limited efforts have been conducted to bring LlaMa-7B to edge devices such as computer systems and smartphones \cite{xu2024fwdllmefficientfedllmusing}. Hence, further research is required to support XLM in edge computing systems. Moreover, the design of scalable and energy-efficient XLM architectures that run on low-capacity hardware without significantly sacrificing performance would be a key enabler of cost-effective XLM-assisted ADS. Such architectures would potentially consider latency and memory optimization, model compression, and knowledge distillation techniques.

\subsection{Advancing Personalized ADS}
Integrating XLMs into ADS marks a paradigm characterized by continuous learning and personalized engagement. Indeed, XLMs can continuously learn from new data and interactions, thus adapting to changing driving patterns, user preferences, and evolving road conditions. This adaptability results in a refined and enhanced performance over time. 
However, real-time personalization in ADS is lacking, which opens numerous opportunities to deploy and validate XLM-assisted personalized AD frameworks. Moreover, the development of XLM-driven virtual assistants that align with drivers’ individual preferences, with safety features, like fatigue detection, and maintenance specifications, can be explored. 

\subsection{Multimodal Retrieval-Augmented Generation Frameworks}
Although retrieval-augmented generation (RAG) techniques, which incorporate relevant external knowledge during text generation resulting in more accurate and contextually relevant outputs, are well addressed in LLMs, Multimodal RAG (MuRAG) is under-explored, especially in ADS \cite{dai2024vistarag,yuan2024rag}. 
Indeed, the latter is expected to combine information retrieval with multimodal data processing and generative capabilities to enhance the AV's understanding and response to complex driving scenarios.  
\subsection{Interplay between Multi-Tasking and Fine-Tuning}
When an XLM in ADS multi-tasks, e.g., used for scene understanding and trajectory prediction, it is hard to fine-tune it efficiently. Novel approaches are needed to coordinate task-specific adjustments while maintaining shared representations among tasks. Also, researchers should explore collaborative sensory modalities in XLM development to obtain a holistic understanding of the driving environment, regarding multi-tasking.   



\subsection{XLM Security}
As XLMs are continuously developing, researchers identify several risks related to them, such as prompt injection, data poisoning, and catastrophic forgetting. These risks present substantial obstacles to the advancement of MLLMs toward their deployment in real-world ADS. Hence, there is an emergency in developing robust security measures to ensure safe and reliable deployment of such advanced models. To do so, techniques such as MLLM-protector, robust training, adversarial defense, and data encryption might be integrated into the framework.

 

    
   

\section{Conclusion}
In this paper, we reviewed XLM techniques and AD frameworks, showing how they both can be integrated. Specifically, we surveyed the most recent LLM, VLM, and MLLM works that are used for AD operations, from the architectural, dataset, and concept perspectives. 
Then, we discussed how they address key ADS challenges, including multi-modal data fusion, safety, reliability, and complex environment understanding.
Among the proposed approaches of XLM for AD, prompt engineering, pre-trained model fine-tuning, RLHF, and GAI methods have been discussed. Through our survey, we emphasize the importance of leveraging XLMs for several AD tasks including planning and control, perception, multi-tasking, and question-answering. 
The role of adequate datasets and simulation tools has been also discussed, and finally, our forward-looking analysis identified open issues and future research directions that would potentially enable practical ADS frameworks assisted by XLMs.

\section{Acknowledgment}
This work is funded in part by the Mitacs Globalink Research Award (GRA) program and in part by the Deanship of Scientific Research at King Abdulaziz University.



\bibliographystyle{ieeetr}
\bibliography{ref}

\begin{thebibliography}{100}

\bibitem{Zhu2019}
L.~{Zhu}, F.~R. {Yu}, Y.~{Wang}, B.~{Ning}, and T.~{Tang}, ``Big data analytics in intelligent transportation systems: A survey,'' {\em IEEE Trans. Intelli. Transport. Syst.}, vol.~20, pp.~383--398, Jan. 2019.

\bibitem{cao2023comprehensive}
Y.~Cao, S.~Li, Y.~Liu, Z.~Yan, Y.~Dai, P.~S. Yu, and L.~Sun, ``A comprehensive survey of {AI}-generated content ({AIGC}): A history of generative {AI} from {GAN} to {ChatGPT},'' {\em arXiv preprint arXiv:2303.04226}, 2023.

\bibitem{ooi2023potential}
K.-B. Ooi, G.~W.-H. Tan, M.~Al-Emran, M.~A. Al-Sharafi, A.~Capatina, A.~Chakraborty, Y.~K. Dwivedi, T.-L. Huang, A.~K. Kar, V.-H. Lee, {\em et~al.}, ``The potential of generative artificial intelligence across disciplines: Perspectives and future directions,'' {\em J. of Comput. Info. Syst.}, pp.~1--32, 2023.

\bibitem{mahor2023iot}
V.~Mahor, R.~Rawat, A.~Kumar, B.~Garg, K.~Pachlasiya, {\em et~al.}, ``{IoT} and artificial intelligence techniques for public safety and security,'' in {\em Smart Urban Comput. Appl.}, pp.~111--126, River Publishers, 2023.

\bibitem{anderljung2023frontier}
M.~Anderljung, J.~Barnhart, J.~Leung, A.~Korinek, C.~O'Keefe, J.~Whittlestone, S.~Avin, M.~Brundage, J.~Bullock, D.~Cass-Beggs, {\em et~al.}, ``Frontier {AI} regulation: Managing emerging risks to public safety,'' {\em arXiv preprint arXiv:2307.03718}, 2023.

\bibitem{chang2024survey}
Y.~Chang, X.~Wang, J.~Wang, Y.~Wu, L.~Yang, K.~Zhu, H.~Chen, X.~Yi, C.~Wang, Y.~Wang, {\em et~al.}, ``A survey on evaluation of large language models,'' {\em ACM Trans. Intelli. Syst. Technol.}, vol.~15, no.~3, pp.~1--45, 2024.

\bibitem{hadi2023survey}
M.~U. Hadi, R.~Qureshi, A.~Shah, M.~Irfan, A.~Zafar, M.~B. Shaikh, N.~Akhtar, J.~Wu, S.~Mirjalili, {\em et~al.}, ``A survey on large language models: Applications, challenges, limitations, and practical usage,'' {\em Authorea Preprints}, 2023.

\bibitem{cui2024survey}
C.~Cui, Y.~Ma, X.~Cao, W.~Ye, Y.~Zhou, K.~Liang, J.~Chen, J.~Lu, Z.~Yang, K.-D. Liao, {\em et~al.}, ``A survey on multimodal large language models for autonomous driving,'' in {\em Proc. IEEE/CVF Winter Conf. Appl. Comput. Vis.}, pp.~958--979, 2024.

\bibitem{ge2023development}
J.~Ge, S.~Sun, J.~Owens, V.~Galvez, O.~Gologorskaya, J.~C. Lai, M.~J. Pletcher, and K.~Lai, ``Development of a liver disease-specific large language model chat interface using retrieval augmented generation,'' {\em medRxiv}, 2023.

\bibitem{chen2023enhancing}
M.~Chen, Z.~Tao, W.~Tang, T.~Qin, R.~Yang, and C.~Zhu, ``Enhancing emergency decision-making with knowledge graphs and large language models,'' {\em arXiv preprint arXiv:2311.08732}, 2023.

\bibitem{wu2024accidentgpt}
K.~Wu, W.~Li, and X.~Xiao, ``{AccidentGPT}: Large multi-modal foundation model for traffic accident analysis,'' {\em arXiv preprint arXiv:2401.03040}, 2024.

\bibitem{qian2024liver}
J.~Qian, Z.~Jin, Q.~Zhang, G.~Cai, and B.~Liu, ``A liver cancer question-answering system based on next-generation intelligence and the large model med-palm 2,'' {\em Int. J. Comp. Sci. Info. Technol.}, vol.~2, no.~1, pp.~28--35, 2024.

\bibitem{singhal2023large}
K.~Singhal, S.~Azizi, T.~Tu, S.~S. Mahdavi, J.~Wei, H.~W. Chung, N.~Scales, A.~Tanwani, H.~Cole-Lewis, S.~Pfohl, {\em et~al.}, ``Large language models encode clinical knowledge,'' {\em Nature}, vol.~620, no.~7972, pp.~172--180, 2023.

\bibitem{ouis2024chestbiox}
M.~Y. Ouis and M.~A. Akhloufi, ``{ChestBioX-Gen}: contextual biomedical report generation from chest {X}-ray images using {BioGPT} and co-attention mechanism,'' {\em Frontiers Imag.}, vol.~3, p.~1373420, 2024.

\bibitem{kafikang2023drug}
M.~KafiKang and A.~Hendawi, ``Drug-drug interaction extraction from biomedical text using relation {BioBERT} with {BLSTM},'' {\em ML \& Knowl. Extract.}, vol.~5, no.~2, pp.~669--683, 2023.

\bibitem{aden2024international}
I.~Aden, C.~H. Child, and C.~C. Reyes-Aldasoro, ``International classification of diseases prediction from {MIMIIC-III} clinical text using pre-trained {ClinicalBERT} and {NLP} deep learning models achieving state of the art,'' {\em Big Data \& Cogn. Comput.}, vol.~8, no.~5, p.~47, 2024.

\bibitem{huang2024leveraging}
Y.~Huang, ``Leveraging large language models for enhanced {NLP} task performance through knowledge distillation and optimized training strategies,'' {\em arXiv preprint arXiv:2402.09282}, 2024.

\bibitem{wang2023review}
J.~Wang, Z.~Liu, L.~Zhao, Z.~Wu, C.~Ma, S.~Yu, H.~Dai, Q.~Yang, Y.~Liu, S.~Zhang, {\em et~al.}, ``Review of large vision models and visual prompt engineering,'' {\em Meta-Radiology}, p.~100047, 2023.

\bibitem{maaz2023video}
M.~Maaz, H.~Rasheed, S.~Khan, and F.~S. Khan, ``{Video-ChatGPT}: Towards detailed video understanding via large vision and language models,'' {\em arXiv preprint arXiv:2306.05424}, 2023.

\bibitem{bai2023sequential}
Y.~Bai, X.~Geng, K.~Mangalam, A.~Bar, A.~Yuille, T.~Darrell, J.~Malik, and A.~A. Efros, ``Sequential modeling enables scalable learning for large vision models,'' in {\em Proc. IEEE/CVF Conf. Comput. Vis. Pattern Recogn. (CVPR)}, 2024.

\bibitem{xu2023lvlm}
P.~Xu, W.~Shao, K.~Zhang, P.~Gao, S.~Liu, M.~Lei, F.~Meng, S.~Huang, Y.~Qiao, and P.~Luo, ``{LVLM-eHUB}: A comprehensive evaluation benchmark for large vision-language models,'' {\em arXiv preprint arXiv:2306.09265}, 2023.

\bibitem{zhao2024evaluating}
Y.~Zhao, T.~Pang, C.~Du, X.~Yang, C.~Li, N.-M.~M. Cheung, and M.~Lin, ``On evaluating adversarial robustness of large vision-language models,'' {\em Adv. Neural Info. Process. Syst.}, vol.~36, 2024.

\bibitem{li2024eyes}
Y.~Li, W.~Tian, Y.~Jiao, J.~Chen, and Y.-G. Jiang, ``Eyes can deceive: Benchmarking counterfactual reasoning abilities of multi-modal large language models,'' {\em arXiv preprint arXiv:2404.12966}, 2024.

\bibitem{zhang2024mm}
D.~Zhang, Y.~Yu, C.~Li, J.~Dong, D.~Su, C.~Chu, and D.~Yu, ``{MM-LLMs}: Recent advances in multimodal large language models,'' {\em arXiv preprint arXiv:2401.13601}, 2024.

\bibitem{yin2023survey}
S.~Yin, C.~Fu, S.~Zhao, K.~Li, X.~Sun, T.~Xu, and E.~Chen, ``A survey on multimodal large language models,'' {\em arXiv preprint arXiv:2306.13549}, 2023.

\bibitem{zhang2024trafficgpt}
S.~Zhang, D.~Fu, W.~Liang, Z.~Zhang, B.~Yu, P.~Cai, and B.~Yao, ``{TrafficGPT}: Viewing, processing and interacting with traffic foundation models,'' {\em Transport Policy}, vol.~150, pp.~95--105, 2024.

\bibitem{zheng2023trafficsafetygpt}
O.~Zheng, M.~Abdel-Aty, D.~Wang, C.~Wang, and S.~Ding, ``{TrafficSafetyGPT}: Tuning a pre-trained large language model to a domain-specific expert in transportation safety,'' {\em arXiv preprint arXiv:2307.15311}, 2023.

\bibitem{gao2024survey}
H.~Gao, Y.~Li, K.~Long, M.~Yang, and Y.~Shen, ``A survey for foundation models in autonomous driving,'' {\em arXiv preprint arXiv:2402.01105}, 2024.

\bibitem{yang2023llm4drive}
Z.~Yang, X.~Jia, H.~Li, and J.~Yan, ``{LLM4Drive}: A survey of large language models for autonomous driving,'' {\em arXiv e-prints}, pp.~arXiv--2311, 2023.

\bibitem{zhou2024vision}
X.~Zhou, M.~Liu, E.~Yurtsever, B.~L. Zagar, W.~Zimmer, H.~Cao, and A.~C. Knoll, ``Vision language models in autonomous driving: A survey and outlook,'' {\em IEEE Trans. Intelli. Veh.}, 2024.

\bibitem{luo2024delving}
S.~Luo, W.~Chen, W.~Tian, R.~Liu, L.~Hou, X.~Zhang, H.~Shen, R.~Wu, S.~Geng, Y.~Zhou, L.~Shao, Y.~Yang, B.~Gao, Q.~Li, and G.~Wu, ``Delving into multi-modal multi-task foundation models for road scene understanding: From learning paradigm perspectives,'' {\em IEEE Trans. Intelli. Veh.}, pp.~1--25, Early Access 2024.

\bibitem{huang2023applications}
Y.~Huang, Y.~Chen, and Z.~Li, ``Applications of large scale foundation models for autonomous driving,'' {\em arXiv preprint arXiv:2311.12144}, 2023.

\bibitem{kim2023completeness}
W.~Kim, J.~H. Kim, Y.~K. Cha, S.~Chong, and T.~J. Kim, ``Completeness of reporting of systematic reviews and meta-analysis of diagnostic test accuracy ({DTA}) of radiological articles based on the {PRISMA-DTA} reporting guideline,'' {\em Acad. Radiology}, vol.~30, no.~2, pp.~258--275, 2023.

\bibitem{janai2020computer}
J.~Janai, F.~G{\"u}ney, A.~Behl, A.~Geiger, {\em et~al.}, ``Computer vision for autonomous vehicles: Problems, datasets and state of the art,'' {\em Foundations \& Trends Comput. Graph. Vis.}, vol.~12, no.~1--3, pp.~1--308, 2020.

\bibitem{parekh2022review}
D.~Parekh, N.~Poddar, A.~Rajpurkar, M.~Chahal, N.~Kumar, G.~P. Joshi, and W.~Cho, ``A review on autonomous vehicles: Progress, methods and challenges,'' {\em Electron.}, vol.~11, no.~14, p.~2162, 2022.

\bibitem{ahangar2021survey}
M.~N. Ahangar, Q.~Z. Ahmed, F.~A. Khan, and M.~Hafeez, ``A survey of autonomous vehicles: Enabling communication technologies and challenges,'' {\em Sensors}, vol.~21, no.~3, p.~706, 2021.

\bibitem{liu2024survey}
M.~Liu, E.~Yurtsever, J.~Fossaert, X.~Zhou, W.~Zimmer, Y.~Cui, B.~L. Zagar, and A.~C. Knoll, ``A survey on autonomous driving datasets: Statistics, annotation quality, and a future outlook,'' {\em IEEE Trans. Intelli. Veh.}, 2024.

\bibitem{brostow2008segmentation}
G.~J. Brostow, J.~Fauqueur, and R.~Cipolla, ``Segmentation and recognition using structure from motion point clouds,'' {\em Proc. Europ. Conf. Comput. Vis.}, pp.~44--57, 2008.

\bibitem{kitti}
A.~Geiger, P.~Lenz, C.~Stiller, and R.~Urtasun, ``{KITTI} vision benchmark suite,'' 2012.
\newblock Accessed: 2024-08-28.

\bibitem{xinyuhuang2020apolloscape}
P.~XinyuHuang, D.~XinjingCheng, Q.~Geng, and R.~Yang, ``The {ApolloScape} open dataset for autonomous driving and its application,'' {\em IEEE Trans. Patt. Anal. Mach. Intelli.}, vol.~42, no.~10, 2020.

\bibitem{apolloscape}
ApolloScape, ``{ApolloScape} dataset,'' 2018.
\newblock Accessed: 2024-08-28.

\bibitem{caesar2020NuScenes}
H.~Caesar, V.~Bankiti, A.~H. Lang, S.~Vora, V.~E. Liong, Q.~Xu, A.~Krishnan, Y.~Pan, G.~Baldan, and O.~Beijbom, ``{nuScenes}: A multimodal dataset for autonomous driving,'' in {\em Proc. IEEE/CVF Conf. Comput. Vis. Patt. Recogn.}, pp.~11621--11631, 2020.

\bibitem{mei2022waymo}
J.~Mei, A.~Z. Zhu, X.~Yan, H.~Yan, S.~Qiao, L.-C. Chen, and H.~Kretzschmar, ``Waymo open dataset: Panoramic video panoptic segmentation,'' in {\em Proc. Europ. Conf. Comput. Vis.}, pp.~53--72, Springer, 2022.

\bibitem{chang2019argoverse}
M.-F. Chang, J.~Lambert, P.~Sangkloy, J.~Singh, S.~Bak, A.~Hartnett, D.~Wang, P.~Carr, S.~Lucey, D.~Ramanan, {\em et~al.}, ``Argoverse: {3D} tracking and forecasting with rich maps,'' in {\em Proc. IEEE/CVF Conf. Comput. Vis. Patt. Recogn.}, pp.~8748--8757, 2019.

\bibitem{yu2020bdd100k}
F.~Yu, H.~Chen, X.~Wang, W.~Xian, Y.~Chen, F.~Liu, V.~Madhavan, and T.~Darrell, ``{BDD100k}: A diverse driving dataset for heterogeneous multitask learning,'' in {\em Proc. IEEE/CVF Conf. Comput. Vis. Patt. Recogn.}, pp.~2636--2645, 2020.

\bibitem{li2022coda}
K.~Li, K.~Chen, H.~Wang, L.~Hong, C.~Ye, J.~Han, Y.~Chen, W.~Zhang, C.~Xu, D.-Y. Yeung, {\em et~al.}, ``{CODA}: A real-world road corner case dataset for object detection in autonomous driving,'' in {\em Europ. Conf. Comput. Vis.}, pp.~406--423, Springer, 2022.

\bibitem{mao2106one}
J.~Mao, M.~Niu, C.~Jiang, H.~Liang, J.~Chen, X.~Liang, Y.~Li, C.~Ye, W.~Zhang, Z.~Li, {\em et~al.}, ``One million scenes for autonomous driving: Once dataset,'' {\em arXiv preprint arXiv:2106.11037}, 2021.

\bibitem{alibeigi2023zenseact}
M.~Alibeigi, W.~Ljungbergh, A.~Tonderski, G.~Hess, A.~Lilja, C.~Lindstr{\"o}m, D.~Motorniuk, J.~Fu, J.~Widahl, and C.~Petersson, ``Zenseact open dataset: A large-scale and diverse multimodal dataset for autonomous driving,'' in {\em Proc. IEEE/CVF Int. Conf. Comput. Vis.}, pp.~20178--20188, 2023.

\bibitem{burnett2023boreas}
K.~Burnett, D.~J. Yoon, Y.~Wu, A.~Z. Li, H.~Zhang, S.~Lu, J.~Qian, W.-K. Tseng, A.~Lambert, K.~Y. Leung, {\em et~al.}, ``Boreas: A multi-season autonomous driving dataset,'' {\em Int. J. Robot. Res.}, vol.~42, no.~1-2, pp.~33--42, 2023.

\bibitem{zhao2023survey}
W.~X. Zhao, K.~Zhou, J.~Li, T.~Tang, X.~Wang, Y.~Hou, Y.~Min, B.~Zhang, J.~Zhang, Z.~Dong, {\em et~al.}, ``A survey of large language models,'' {\em arXiv preprint arXiv:2303.18223}, 2023.

\bibitem{raiaan2024review}
M.~A.~K. Raiaan, M.~S.~H. Mukta, K.~Fatema, N.~M. Fahad, S.~Sakib, M.~M.~J. Mim, J.~Ahmad, M.~E. Ali, and S.~Azam, ``A review on large language models: Architectures, applications, taxonomies, open issues and challenges,'' {\em IEEE Access}, 2024.

\bibitem{vaswani2017attention}
A.~Vaswani, N.~Shazeer, N.~Parmar, J.~Uszkoreit, L.~Jones, A.~N. Gomez, {\L}.~Kaiser, and I.~Polosukhin, ``Attention is all you need,'' {\em Adv. Neural Info. Process. Syst.}, vol.~30, 2017.

\bibitem{marvin2023prompt}
G.~Marvin, N.~Hellen, D.~Jjingo, and J.~Nakatumba-Nabende, ``Prompt engineering in large language models,'' in {\em Proc. Int. Conf. Data Intelli. Cogn. Informat.}, pp.~387--402, Springer, 2023.

\bibitem{han2024parameter}
Z.~Han, C.~Gao, J.~Liu, S.~Q. Zhang, {\em et~al.}, ``Parameter-efficient fine-tuning for large models: A comprehensive survey,'' {\em arXiv preprint arXiv:2403.14608}, 2024.

\bibitem{sun2023evaluating}
J.~Sun, C.~Shaib, and B.~C. Wallace, ``Evaluating the zero-shot robustness of instruction-tuned language models,'' {\em arXiv preprint arXiv:2306.11270}, 2023.

\bibitem{chan2024dense}
A.~J. Chan, H.~Sun, S.~Holt, and M.~van~der Schaar, ``Dense reward for free in reinforcement learning from human feedback,'' {\em arXiv preprint arXiv:2402.00782}, 2024.

\bibitem{dosovitskiy2020image}
A.~Dosovitskiy, L.~Beyer, A.~Kolesnikov, D.~Weissenborn, X.~Zhai, T.~Unterthiner, M.~Dehghani, M.~Minderer, G.~Heigold, S.~Gelly, {\em et~al.}, ``An image is worth 16x16 words: Transformers for image recognition at scale,'' {\em arXiv preprint arXiv:2010.11929}, 2020.

\bibitem{touvron2021training}
H.~Touvron, M.~Cord, M.~Douze, F.~Massa, A.~Sablayrolles, and H.~J{\'e}gou, ``Training data-efficient image transformers \& distillation through attention,'' in {\em Proc. Int. Conf. Mach. Learn.}, pp.~10347--10357, PMLR, 2021.

\bibitem{yuan2021tokens}
L.~Yuan, Y.~Chen, T.~Wang, W.~Yu, Y.~Shi, Z.-H. Jiang, F.~E. Tay, J.~Feng, and S.~Yan, ``Tokens-to-token {ViT}: Training vision transformers from scratch on imagenet,'' in {\em Proc. IEEE/CVF Int. Conf. Comput. Vis.}, pp.~558--567, 2021.

\bibitem{liu2021swin}
Z.~Liu, Y.~Lin, Y.~Cao, H.~Hu, Y.~Wei, Z.~Zhang, S.~Lin, and B.~Guo, ``{Swin Transformer}: Hierarchical vision transformer using shifted windows,'' in {\em Proc. IEEE/CVF Int. Conf. Comput. Vis.}, pp.~10012--10022, 2021.

\bibitem{he2022masked}
K.~He, X.~Chen, S.~Xie, Y.~Li, P.~Doll{\'a}r, and R.~Girshick, ``Masked autoencoders are scalable vision learners,'' in {\em Proc. IEEE/CVF Conf. Comput. Vis. Patt. Recogn.}, pp.~16000--16009, 2022.

\bibitem{li2022exploring}
Y.~Li, H.~Mao, R.~Girshick, and K.~He, ``Exploring plain vision transformer backbones for object detection,'' in {\em Europ. Conf. Comput. Vis.}, pp.~280--296, Springer, 2022.

\bibitem{wang2021pyramid}
W.~Wang, E.~Xie, X.~Li, D.-P. Fan, K.~Song, D.~Liang, T.~Lu, P.~Luo, and L.~Shao, ``Pyramid vision transformer: A versatile backbone for dense prediction without convolutions,'' in {\em Proc. IEEE/CVF Int. Conf. Comput. Vis.}, pp.~568--578, 2021.

\bibitem{chen2021crossvit}
C.-F.~R. Chen, Q.~Fan, and R.~Panda, ``{CrossViT}: Cross-attention multi-scale vision transformer for image classification,'' in {\em Proc. IEEE/CVF Int. Conf. Comput. Vis.}, pp.~357--366, 2021.

\bibitem{meng2022adavit}
L.~Meng, H.~Li, B.-C. Chen, S.~Lan, Z.~Wu, Y.-G. Jiang, and S.-N. Lim, ``{AdaViT}: Adaptive vision transformers for efficient image recognition,'' in {\em Proc. IEEE/CVF Conf. Comput. Vis. Patt. Recogn.}, pp.~12309--12318, 2022.

\bibitem{carion2020end}
N.~Carion, F.~Massa, G.~Synnaeve, N.~Usunier, A.~Kirillov, and S.~Zagoruyko, ``End-to-end object detection with transformers,'' in {\em Europ. Conf. Comput. Vis.}, pp.~213--229, Springer, 2020.

\bibitem{xie2021segformer}
E.~Xie, W.~Wang, Z.~Yu, A.~Anandkumar, J.~M. Alvarez, and P.~Luo, ``{SegFormer}: Simple and efficient design for semantic segmentation with transformers,'' {\em Adv. Neural Info. Process. Syst.}, vol.~34, pp.~12077--12090, 2021.

\bibitem{oquab2023dinov2}
M.~Oquab, T.~Darcet, T.~Moutakanni, H.~Vo, M.~Szafraniec, V.~Khalidov, P.~Fernandez, D.~Haziza, F.~Massa, A.~El-Nouby, {\em et~al.}, ``Dinov2: Learning robust visual features without supervision,'' {\em arXiv preprint arXiv:2304.07193}, 2023.

\bibitem{bao2021beit}
H.~Bao, L.~Dong, S.~Piao, and F.~Wei, ``Beit: {BERT} pre-training of image transformers,'' {\em arXiv preprint arXiv:2106.08254}, 2021.

\bibitem{radford2021learning}
A.~Radford, J.~W. Kim, C.~Hallacy, A.~Ramesh, G.~Goh, S.~Agarwal, G.~Sastry, A.~Askell, P.~Mishkin, J.~Clark, {\em et~al.}, ``Learning transferable visual models from natural language supervision,'' in {\em Proc. Int. Conf. Mach. Learn.}, pp.~8748--8763, PMLR, 2021.

\bibitem{ramesh2021zero}
A.~Ramesh, M.~Pavlov, G.~Goh, S.~Gray, C.~Voss, A.~Radford, M.~Chen, and I.~Sutskever, ``Zero-shot text-to-image generation,'' in {\em Proc. Int. Conf. Mach. Learn.}, pp.~8821--8831, Pmlr, 2021.

\bibitem{ramesh2022hierarchical}
A.~Ramesh, P.~Dhariwal, A.~Nichol, C.~Chu, and M.~Chen, ``Hierarchical text-conditional image generation with clip latents,'' {\em arXiv preprint arXiv:2204.06125}, vol.~1, no.~2, p.~3, 2022.

\bibitem{chen2020uniter}
Y.-C. Chen, L.~Li, L.~Yu, A.~El~Kholy, F.~Ahmed, Z.~Gan, Y.~Cheng, and J.~Liu, ``{UNITER}: Universal image-text representation learning,'' in {\em Europ. Conf. Comput. Vis.}, pp.~104--120, Springer, 2020.

\bibitem{wang2021simvlm}
Z.~Wang, J.~Yu, A.~W. Yu, Z.~Dai, Y.~Tsvetkov, and Y.~Cao, ``{SimVLM}: Simple visual language model pretraining with weak supervision,'' {\em arXiv preprint arXiv:2108.10904}, 2021.

\bibitem{wang2023image}
W.~Wang, H.~Bao, L.~Dong, J.~Bjorck, Z.~Peng, Q.~Liu, K.~Aggarwal, O.~K. Mohammed, S.~Singhal, S.~Som, {\em et~al.}, ``Image as a foreign language: Beit pretraining for vision and vision-language tasks,'' in {\em Proc. IEEE/CVF Conf. Comput. Vis. Patt. Recogn.}, pp.~19175--19186, 2023.

\bibitem{driess2023palm}
D.~Driess, F.~Xia, M.~S. Sajjadi, C.~Lynch, A.~Chowdhery, B.~Ichter, A.~Wahid, J.~Tompson, Q.~Vuong, T.~Yu, {\em et~al.}, ``{Palm-E}: An embodied multimodal language model,'' in {\em Proc. Int. Conf. ML (ICML)}, 2023.

\bibitem{huang2024language}
S.~Huang, L.~Dong, W.~Wang, Y.~Hao, S.~Singhal, S.~Ma, T.~Lv, L.~Cui, O.~K. Mohammed, B.~Patra, {\em et~al.}, ``Language is not all you need: Aligning perception with language models,'' {\em Adv. Neural Info. Process. Syst.}, vol.~36, 2024.

\bibitem{xu2024survey}
M.~Xu, W.~Yin, D.~Cai, R.~Yi, D.~Xu, Q.~Wang, B.~Wu, Y.~Zhao, C.~Yang, S.~Wang, {\em et~al.}, ``A survey of resource-efficient {LLM} and multimodal foundation models,'' {\em arXiv preprint arXiv:2401.08092}, 2024.

\bibitem{wang2024exploring}
Y.~Wang, W.~Chen, X.~Han, X.~Lin, H.~Zhao, Y.~Liu, B.~Zhai, J.~Yuan, Q.~You, and H.~Yang, ``Exploring the reasoning abilities of multimodal large language models ({MLLM}s): A comprehensive survey on emerging trends in multimodal reasoning,'' {\em arXiv preprint arXiv:2401.06805}, 2024.

\bibitem{wu2023multimodal}
J.~Wu, W.~Gan, Z.~Chen, S.~Wan, and S.~Y. Philip, ``Multimodal large language models: A survey,'' in {\em Proc. IEEE Int. Conf. Big Data (BigData)}, pp.~2247--2256, IEEE, 2023.

\bibitem{xiang2023multi}
C.~Xiang, C.~Feng, X.~Xie, B.~Shi, H.~Lu, Y.~Lv, M.~Yang, and Z.~Niu, ``Multi-sensor fusion and cooperative perception for autonomous driving: A review,'' {\em IEEE Intelli. Transport. Syst. Mag.}, 2023.

\bibitem{singh2023transformer}
A.~Singh, ``Transformer-based sensor fusion for autonomous driving: A survey,'' in {\em Proc. IEEE/CVF Int. Conf. Comput. Vis.}, pp.~3312--3317, 2023.

\bibitem{zhao2023potential}
X.~Zhao, Y.~Fang, H.~Min, X.~Wu, W.~Wang, and R.~Teixeira, ``Potential sources of sensor data anomalies for autonomous vehicles: An overview from road vehicle safety perspective,'' {\em Expert Syst. Appl.}, p.~121358, 2023.

\bibitem{hasanujjaman2023sensor}
M.~Hasanujjaman, M.~Z. Chowdhury, and Y.~M. Jang, ``Sensor fusion in autonomous vehicle with traffic surveillance camera system: detection, localization, and {AI} networking,'' {\em Sensors}, vol.~23, no.~6, p.~3335, 2023.

\bibitem{choi2023semantics}
H.-S. Choi, J.~Jeong, Y.~H. Cho, K.-J. Yoon, and J.-H. Kim, ``Semantics-guided transformer-based sensor fusion for improved waypoint prediction,'' {\em arXiv preprint arXiv:2308.02126}, 2023.

\bibitem{nouri2024engineering}
A.~Nouri, B.~Cabrero-Daniel, F.~Törner, H.~Sivencrona, and C.~Berger, ``Engineering safety requirements for autonomous driving with large language models,'' in {\em Proc. IEEE Int. Requirements Engineer. Conf. (RE)}, pp.~218--228, 2024.

\bibitem{wang2023empowering}
Y.~Wang, R.~Jiao, C.~Lang, S.~S. Zhan, C.~Huang, Z.~Wang, Z.~Yang, and Q.~Zhu, ``Empowering autonomous driving with large language models: A safety perspective,'' {\em arXiv preprint arXiv:2312.00812}, 2023.

\bibitem{wang2023accidentgpt}
L.~Wang, H.~Jiang, P.~Cai, D.~Fu, T.~Wang, Z.~Cui, Y.~Ren, H.~Yu, X.~Wang, and Y.~Wang, ``Accident{GPT}: Accident analysis and prevention from {V2X} environmental perception with multi-modal large model,'' {\em arXiv preprint arXiv:2312.13156}, 2023.

\bibitem{aldeenwip}
M.~Aldeen, P.~MohajerAnsari, J.~Ma, M.~Chowdhury, L.~Cheng, and M.~D. Pes{\'e}, ``{WIP}: A first look at employing large multimodal models against autonomous vehicle attacks,'' {\em Proc. Symp. Veh. Secu. Priv. (VehicleSec)}, 2024.

\bibitem{cui2023human}
C.~Cui, Y.~Ma, X.~Cao, W.~Ye, and Z.~Wang, ``Human-autonomy teaming on autonomous vehicles with large language model-enabled human digital twins,'' in {\em Proc. IEEE/ACM Symp. Edge Comput. (SEC)}, pp.~319--324, IEEE, 2023.

\bibitem{yang2024human}
Y.~Yang, Q.~Zhang, C.~Li, D.~S. Marta, N.~Batool, and J.~Folkesson, ``Human-centric autonomous systems with {LLM}s for user command reasoning,'' in {\em Proc. IEEE/CVF Winter Conf. Appl. Comput. Vis.}, pp.~988--994, 2024.

\bibitem{xu2023drivegpt4}
Z.~Xu, Y.~Zhang, E.~Xie, Z.~Zhao, Y.~Guo, K.-Y.~K. Wong, Z.~Li, and H.~Zhao, ``{DriveGPT4}: Interpretable end-to-end autonomous driving via large language model,'' {\em IEEE Robotics Autom. Lett.}, vol.~9, no.~10, pp.~8186--8193, 2024.

\bibitem{tanahashi2023evaluation}
K.~Tanahashi, Y.~Inoue, Y.~Yamaguchi, H.~Yaginuma, D.~Shiotsuka, H.~Shimatani, K.~Iwamasa, Y.~Inoue, T.~Yamaguchi, K.~Igari, {\em et~al.}, ``Evaluation of large language models for decision making in autonomous driving,'' {\em arXiv preprint arXiv:2312.06351}, 2023.

\bibitem{zhou2024context}
Z.~Zhou, J.~Zhang, J.~Zhang, B.~Wang, T.~Shi, and A.~Khamis, ``In-context learning for automated driving scenarios,'' {\em arXiv preprint arXiv:2405.04135}, 2024.

\bibitem{wen2023dilu}
L.~Wen, D.~Fu, X.~Li, X.~Cai, T.~Ma, P.~Cai, M.~Dou, B.~Shi, L.~He, and Y.~Qiao, ``{DiLu}: A knowledge-driven approach to autonomous driving with large language models,'' {\em arXiv preprint arXiv:2309.16292}, 2023.

\bibitem{sha2023languagempc}
H.~Sha, Y.~Mu, Y.~Jiang, L.~Chen, C.~Xu, P.~Luo, S.~E. Li, M.~Tomizuka, W.~Zhan, and M.~Ding, ``Language{MPC}: Large language models as decision makers for autonomous driving,'' {\em arXiv preprint arXiv:2310.03026}, 2023.

\bibitem{azarafza2024hybrid}
M.~Azarafza, M.~Nayyeri, C.~Steinmetz, S.~Staab, and A.~Rettberg, ``Hybrid reasoning based on large language models for autonomous car driving,'' {\em arXiv preprint arXiv:2402.13602}, 2024.

\bibitem{jin2023surrealdriver}
Y.~Jin, X.~Shen, H.~Peng, X.~Liu, J.~Qin, J.~Li, J.~Xie, P.~Gao, G.~Zhou, and J.~Gong, ``{SurrealDriver}: Designing generative driver agent simulation framework in urban contexts based on large language model,'' {\em arXiv preprint arXiv:2309.13193}, 2023.

\bibitem{miceli2023dialogue}
A.~V. Miceli-Barone, A.~Lascarides, and C.~Innes, ``Dialogue-based generation of self-driving simulation scenarios using large language models,'' {\em arXiv preprint arXiv:2310.17372}, 2023.

\bibitem{chen2023driving}
L.~Chen, O.~Sinavski, J.~Hünermann, A.~Karnsund, A.~J. Willmott, D.~Birch, D.~Maund, and J.~Shotton, ``Driving with {LLMs}: Fusing object-level vector modality for explainable autonomous driving,'' in {\em Proc. IEEE Int. Conf. Robot. Automa. (ICRA)}, pp.~14093--14100, 2024.

\bibitem{wang2024drivecot}
T.~Wang, E.~Xie, R.~Chu, Z.~Li, and P.~Luo, ``{DriveCoT}: Integrating chain-of-thought reasoning with end-to-end driving,'' {\em arXiv preprint arXiv:2403.16996}, 2024.

\bibitem{peng2024lc}
M.~Peng, X.~Guo, X.~Chen, M.~Zhu, K.~Chen, X.~Wang, Y.~Wang, {\em et~al.}, ``{LC-LLM}: Explainable lane-change intention and trajectory predictions with large language models,'' {\em arXiv preprint arXiv:2403.18344}, 2024.

\bibitem{yang2024driving}
R.~Yang, X.~Zhang, A.~Fernandez-Laaksonen, X.~Ding, and J.~Gong, ``Driving style alignment for {LLM}-powered driver agent,'' {\em arXiv preprint arXiv:2403.11368}, 2024.

\bibitem{tian2024enhancing}
H.~Tian, K.~Reddy, Y.~Feng, M.~Quddus, Y.~Demiris, and P.~Angeloudis, ``Enhancing autonomous vehicle training with language model integration and critical scenario generation,'' {\em arXiv preprint arXiv:2404.08570}, 2024.

\bibitem{zhao2024drivedreamer}
G.~Zhao, X.~Wang, Z.~Zhu, X.~Chen, G.~Huang, X.~Bao, and X.~Wang, ``{DriveDreamer-2}: {LLM}-enhanced world models for diverse driving video generation,'' {\em arXiv preprint arXiv:2403.06845}, 2024.

\bibitem{wang2023drivedreamer}
X.~Wang, Z.~Zhu, G.~Huang, X.~Chen, and J.~Lu, ``{DriveDreamer}: Towards real-world-driven world models for autonomous driving,'' {\em arXiv preprint arXiv:2309.09777}, 2023.

\bibitem{guo2024co}
Z.~Guo, A.~Lykov, Z.~Yagudin, M.~Konenkov, and D.~Tsetserukou, ``Co-driver: {VLM}-based autonomous driving assistant with human-like behavior and understanding for complex road scenes,'' {\em arXiv preprint arXiv:2405.05885}, 2024.

\bibitem{tian2024drivevlm}
X.~Tian, J.~Gu, B.~Li, Y.~Liu, C.~Hu, Y.~Wang, K.~Zhan, P.~Jia, X.~Lang, and H.~Zhao, ``{DriveVLM}: The convergence of autonomous driving and large vision-language models,'' {\em arXiv preprint arXiv:2402.12289}, 2024.

\bibitem{li2024automated}
Y.~Li, W.~Zhang, K.~Chen, Y.~Liu, P.~Li, R.~Gao, L.~Hong, M.~Tian, X.~Zhao, Z.~Li, {\em et~al.}, ``Automated evaluation of large vision-language models on self-driving corner cases,'' {\em arXiv preprint arXiv:2404.10595}, 2024.

\bibitem{kou2024pfedlvm}
W.-B. Kou, Q.~Lin, M.~Tang, S.~Xu, R.~Ye, Y.~Leng, S.~Wang, Z.~Chen, G.~Zhu, and Y.-C. Wu, ``{pFedLVM}: A large vision model ({LVM})-driven and latent feature-based personalized federated learning framework in autonomous driving,'' {\em arXiv preprint arXiv:2405.04146}, 2024.

\bibitem{gopalkrishnan2024multi}
A.~Gopalkrishnan, R.~Greer, and M.~Trivedi, ``Multi-frame, lightweight \& efficient vision-language models for question answering in autonomous driving,'' {\em arXiv preprint arXiv:2403.19838}, 2024.

\bibitem{wu2023language}
D.~Wu, W.~Han, T.~Wang, Y.~Liu, X.~Zhang, and J.~Shen, ``Language prompt for autonomous driving,'' {\em arXiv preprint arXiv:2309.04379}, 2023.

\bibitem{ding2023hilm}
X.~Ding, J.~Han, H.~Xu, W.~Zhang, and X.~Li, ``{HiLM-D}: Towards high-resolution understanding in multimodal large language models for autonomous driving,'' {\em arXiv preprint arXiv:2309.05186}, 2023.

\bibitem{shukor2023ep}
M.~Shukor, C.~Dancette, and M.~Cord, ``{EP-ALM}: Efficient perceptual augmentation of language models,'' in {\em Proc. IEEE/CVF Int. Conf. Comput. Vis.}, pp.~22056--22069, 2023.

\bibitem{zhang2023video}
H.~Zhang, X.~Li, and L.~Bing, ``{Video-Llama}: An instruction-tuned audio-visual language model for video understanding,'' in {\em Proc. Conf. Empir. Methods Nat. Lang. Process.: Syst. Demo.}, 2023.

\bibitem{yuan2024rag}
J.~Yuan, S.~Sun, D.~Omeiza, B.~Zhao, P.~Newman, L.~Kunze, and M.~Gadd, ``{RAG-Driver}: Generalisable driving explanations with retrieval-augmented in-context learning in multi-modal large language model,'' {\em arXiv preprint arXiv:2402.10828}, 2024.

\bibitem{wang2023drivemlm}
W.~Wang, J.~Xie, C.~Hu, H.~Zou, J.~Fan, W.~Tong, Y.~Wen, S.~Wu, H.~Deng, Z.~Li, {\em et~al.}, ``{DriveMLM}: Aligning multi-modal large language models with behavioral planning states for autonomous driving,'' {\em arXiv preprint arXiv:2312.09245}, 2023.

\bibitem{han2024dme}
W.~Han, D.~Guo, C.-Z. Xu, and J.~Shen, ``{DME-Driver}: Integrating human decision logic and {3D} scene perception in autonomous driving,'' {\em arXiv preprint arXiv:2401.03641}, 2024.

\bibitem{huangdrivlme}
Y.~Huang, J.~Sansom, Z.~Ma, F.~Gervits, and J.~Chai, ``{DriVLMe}: Exploring foundation models as autonomous driving agents that perceive, communicate, and navigate,'' in {\em Proc. Vis. \& Lang. for Autonom. Driv. Robot. Wrkshp.}, 2024.

\bibitem{liao2024vlm2scene}
G.~Liao, J.~Li, and X.~Ye, ``{VLM2Scene}: Self-supervised image-text-{LiDAR} learning with foundation models for autonomous driving scene understanding,'' in {\em Proc. AAAI Conf. Artifi. Intelli.}, pp.~3351--3359, 2024.

\bibitem{liang2024aide}
M.~Liang, J.-C. Su, S.~Schulter, S.~Garg, S.~Zhao, Y.~Wu, and M.~Chandraker, ``{AIDE}: An automatic data engine for object detection in autonomous driving,'' in {\em Proc. IEEE/CVF Conf. Comput. Vis. Pattern Recogn. (CVPR)}, 2024.

\bibitem{ding2024holistic}
X.~Ding, J.~Han, H.~Xu, X.~Liang, W.~Zhang, and X.~Li, ``Holistic autonomous driving understanding by bird's-eye-view injected multi-modal large models,'' in {\em Proc. IEEE/CVF Conf. Comput. Vis. Pattern Recogn. (CVPR)}, 2024.

\bibitem{wang2024omnidrive}
S.~Wang, Z.~Yu, X.~Jiang, S.~Lan, M.~Shi, N.~Chang, J.~Kautz, Y.~Li, and J.~M. Alvarez, ``{OmniDrive}: A holistic {LLM}-agent framework for autonomous driving with {3D} perception, reasoning and planning,'' {\em arXiv preprint arXiv:2405.01533}, 2024.

\bibitem{mao2023language}
J.~Mao, J.~Ye, Y.~Qian, M.~Pavone, and Y.~Wang, ``A language agent for autonomous driving,'' {\em arXiv preprint arXiv:2311.10813}, 2023.

\bibitem{rayfeedback}
J.~Z. Z. H.~A. Ray and E.~Ohn-Bar, ``Feedback-guided autonomous driving,'' {\em Proc. IEEE/CVF Conf. Comput. Vis. Patt. Recogn. (CVPR)}, pp.~15000--15011, 2024.

\bibitem{jia2023adriver}
F.~Jia, W.~Mao, Y.~Liu, Y.~Zhao, Y.~Wen, C.~Zhang, X.~Zhang, and T.~Wang, ``{Adriver-I}: A general world model for autonomous driving,'' {\em arXiv preprint arXiv:2311.13549}, 2023.

\bibitem{wei2024editable}
Y.~Wei, Z.~Wang, Y.~Lu, C.~Xu, C.~Liu, H.~Zhao, S.~Chen, and Y.~Wang, ``Editable scene simulation for autonomous driving via collaborative {LLM}-agents,'' in {\em Proc. IEEE/CVF Conf. Comput. Vis. Pattern Recogn. (CVPR)}, 2024.

\bibitem{malla2023drama}
S.~Malla, C.~Choi, I.~Dwivedi, J.~H. Choi, and J.~Li, ``Drama: Joint risk localization and captioning in driving,'' in {\em Proc. IEEE/CVF Winter Conf. Appl. Comput. Vis.}, pp.~1043--1052, 2023.

\bibitem{inoue2024NuScenes}
Y.~Inoue, Y.~Yada, K.~Tanahashi, and Y.~Yamaguchi, ``nuscenes-mqa: Integrated evaluation of captions and qa for autonomous driving datasets using markup annotations,'' in {\em Proc. IEEE/CVF Winter Conf. Appl. Comput. Vis.}, pp.~930--938, 2024.

\bibitem{fu2024limsim++}
D.~Fu, W.~Lei, L.~Wen, P.~Cai, S.~Mao, M.~Dou, B.~Shi, and Y.~Qiao, ``{LimSim}++: A closed-loop platform for deploying multimodal {LLMs} in autonomous driving,'' {\em arXiv preprint arXiv:2402.01246}, 2024.

\bibitem{bai2024hallucination}
Z.~Bai, P.~Wang, T.~Xiao, T.~He, Z.~Han, Z.~Zhang, and M.~Z. Shou, ``Hallucination of multimodal large language models: A survey,'' {\em arXiv preprint arXiv:2404.18930}, 2024.

\bibitem{xu2024fwdllmefficientfedllmusing}
M.~Xu, D.~Cai, Y.~Wu, X.~Li, and S.~Wang, ``{FwdLLM}: Efficient {FedLLM} using forward gradient,'' {\em arXiv}, 2024.

\bibitem{dai2024vistarag}
X.~Dai, C.~Guo, Y.~Tang, H.~Li, Y.~Wang, J.~Huang, Y.~Tian, X.~Xia, Y.~Lv, and F.-Y. Wang, ``{VistaRAG}: Toward safe and trustworthy autonomous driving through retrieval-augmented generation,'' {\em IEEE Trans. Intelli. Veh.}, 2024.

\end{thebibliography}

\begin{IEEEbiography}[{\includegraphics[width=1in,height=1.25in,clip,keepaspectratio]{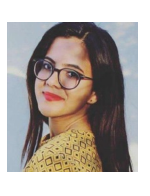}}]{Sonda Fourati }is a student in Computer Systems Engineering at the Mediterranean Institute of Technology (MedTech) in Tunis, Tunisia. With a passion for cutting-edge technology, Sonda Fourati's research focuses on autonomous driving systems and the integration of Multimodal Large Language Models (MLLM) to enhance decision-making in such systems. In addition to a strong technical foundation in systems engineering, Ms. Fourati has developed expertise in artificial intelligence, machine learning, and computer vision, contributing to advancements in the field of autonomous vehicle technology.
\end{IEEEbiography}

\begin{IEEEbiography}[{\includegraphics[width=1in,height=1.25in,clip,keepaspectratio]{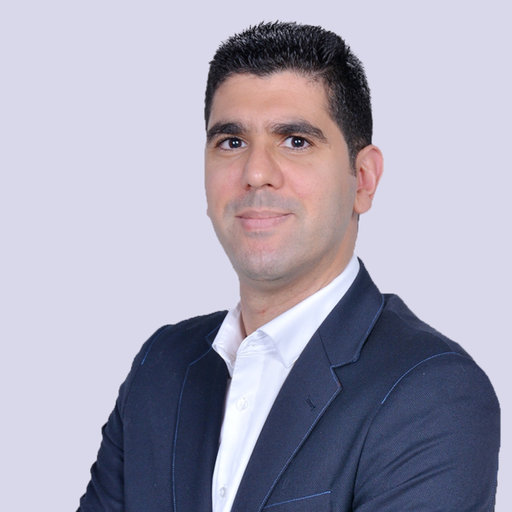}}]{Wael Jaafar }  (S’08, M’14, SM’20) is an Associate Professor at the Software and IT Engineering Department of École de Technologie Supérieure (ÉTS), University of Quebec, Montreal, Canada since September 2022. He holds Master and PhD degrees from Polytechnique Montreal, Canada. Between 2019 and 2022, Dr. Jaafar was with the Systems and Computer Engineering Department of Carleton University as an NSERC Postdoctoral Fellow. From 2014 to 2018, he has pursued a career in the telecommunications industry, where he has been involved in designing telecom solutions for projects across Canada and abroad.  
During his career, Dr. Jaafar was a visiting researcher at Khalifa University, Abu Dhabi, UAE in 2019, Keio University, Japan in 2013, and UQAM, Canada in 2007. He is the recipient of several prestigious grants including NSERC Alexandre-Graham Bell scholarship, FRQNT internship scholarship, and best paper awards at IEEE ICC 2021 and ISCC 2023. His current research interests include wireless communications, integrated terrestrial and non-terrestrial networks, resource allocation, edge caching and computing, and machine learning for communications and networks.
\end{IEEEbiography}

\begin{IEEEbiography}[{\includegraphics[width=1in,height=1.25in,clip,keepaspectratio]{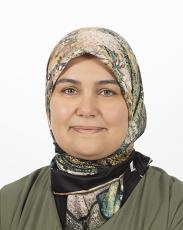}}]{Noura Baccar}\; holds a Ph.D. in Communication Systems from the National Engineering School of Tunis (ENIT) within the research Laboratory Innov’COM at the High School of Telecommunication of Tunis (Sup’Com), Tunisia. She is currently an Assistant Professor and Director of the Computer Systems department at the Mediterranean Institute of Technology (MedTech), South Mediterranean University, Tunisia. During her Ph.D. studies, She Received full scholarship funding ``Mobidoc'' financed by the European Union (EU) and supported by Cynapsys IT company. Dr. Baccar has been a researcher within its R\&D department where she was conducting and supervising many hands-on projects on localization in wireless sensor networks. 
Dr. Baccar is also a Reviewer and Member of the IEEE community. She is also a member of the research group ``Innovation on Advanced Computer technologies'' (I-ACT) in Medtech. Her research interests include electronics, digital systems design, embedded and communication systems, fuzzy logic-based modeling, and artificial intelligence.
\end{IEEEbiography}

\begin{IEEEbiography}[{\includegraphics[width=1in,height=1.25in,clip,keepaspectratio]{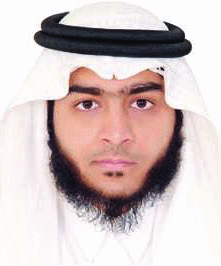}}]{Safwan Alfattani} \; (Member, IEEE) 
is an Assistant Professor at King AbdulAziz University (KAU) in Rabigh, Saudi Arabia, within the Faculty of Engineering’s Electrical Engineering Department. He earned both his Ph.D. and M.Sc. from the University of Ottawa, Canada. His research interests include wireless communications, non-terrestrial networks, Internet of Things (IoT) networks, and reconfigurable intelligent surfaces (RIS). Dr. Alfattani has made significant contributions to these fields through various publications, such as his works on high altitude platform station (HAPS) networks, aerial platforms with reconfigurable smart surfaces for 5G and beyond, and the future applications of LiFi technology. He is recognized for his advancements in wireless communication technologies and his active role in academic research, rewarding him with the grand prize of the IEEE Future Networks competition on NTNs for B5G and 6G, in 2022.

\end{IEEEbiography}

\end{document}